\documentclass[11pt,a4paper]{article}
\usepackage{jheppub}
\usepackage{amsfonts,amsmath,url, xcolor}
\usepackage{eurosym}
\usepackage{braket}
\usepackage{graphicx}
\usepackage[utf8]{inputenc}
\usepackage{BOONDOX-cal}
\newtheorem{theorem}{Theorem}

\usepackage{tikz}
\usepackage{caption}
\usepackage{graphicx}
\graphicspath{ {./images/} }

\begin{document}

\title{
3-dimensional $\Lambda$-BMS Symmetry and its Deformations }

\author[a]{Andrzej Borowiec}
\emailAdd{andrzej.borowiec@uwr.edu.pl,jerzy.kowalski-glikman@uwr.edu.pl,\\unger.josua@uwr.edu.pl }
\author[a,b]{Jerzy Kowalski-Glikman}
\author[a]{Josua Unger}

\affiliation[a]{Institute for Theoretical Physics, University of Wroc\l{}aw, pl.\ M.\ Borna 9, 50-204 Wroc\l{}aw, Poland}
\affiliation[b]{National Centre for Nuclear Research, Pasteura 7, 02-093 Warsaw, Poland}

\abstract{
In this paper we study quantum group deformations of the infinite dimensional symmetry algebra of asymptotically AdS spacetimes in three dimensions. Building on previous results in the finite dimensional subalgebras we classify all possible Lie bialgebra structures  and for selected examples we explicitely construct the related Hopf algebras.
 Using cohomological arguments we show that this construction can always be performed by a so-called twist deformation. The resulting structures can be compared to the well-known $\kappa$-Poincar\'e Hopf algebras constructed on the finite dimensional Poincar\'e or (anti) de Sitter algebra. The dual $\kappa$ Minkowski spacetime is supposed to describe a specific non-commutative geometry. Importantly, we find that some incarnations of the $\kappa$-Poincar\'e can not be extended consistently to the infinite dimensional algebras. Furthermore, certain deformations can have  potential physical applications if subalgebras are considered.  Since the conserved charges associated with asymptotic symmetries in 3-dimensional form a centrally extended algebra we also discuss briefly deformations of such algebras. The presence of the full symmetry algebra might have observable consequences that could be used to rule out these deformations. }
%\arxivnumber{}

\date{\today}

\maketitle

%\ab{This is an example comment!}\\
%\jkg{Second comment!}\\
%\ju{Third comment}

\section{Introduction}

Gravity in 3 dimensions \cite{ch:I-1Staruszkiewicz:1963zza,Deser:1983tn,Deser:1983nh}, see \cite{ch:I-1Carlip:1998uc} for review, is a remarkably rich and interesting theory. It remained relatively obscured for many years, but from the seminal paper of Witten \cite{ch:I-2Witten:1988hc} (see also \cite{ch:I-2Achucarro:1987vz}) it became one of the most studied theory in theoretical physics.

There are many reasons for that. Gravity in 3 dimensions is a topological field theory with no local degrees of freedom, which makes the quantum theory exactly soluble, so it can serve as a toy model of quantum gravity. As it was shown in \cite{Brown:1986nw} in the case of negative cosmological constant this theory possesses an asymptotic Virasoro symmetry. This result was a precursor of AdS/CFT \cite{Maldacena:1997re} and the AdS$_3$/CFT$_2$ correspondence is actively and intensively investigated \cite{Kraus:2006wn}, \cite{Freidel:2008sh}. Second, in spite of having no local dynamical degrees of freedom, 3 dimensional gravity with negative cosmological constant admits black hole solution \cite{Banados:1992wn,Banados:1992gq}, which makes 3-dimensional gravity a nice toy model for studying Hawking radiation. See \cite{Carlip:2005zn} for a review of these aspects of the theory.

Another interesting property of 3-dimensional gravity is the fact that it provides a model of emergence of quantum group symmetries as physical symmetries of quantized gravitating systems. Envisioned in \cite{Witten:1988hf} and based on mathematical works \cite{Fock:1998nu,Alekseev:1993qs,Alekseev:1993rj} this idea was developed further, among others, in canonical formulation in \cite{Bais:2002ye,Meusburger:2003ta,Meusburger:2003hc,Meusburger:2007ad}, and for gravitational path integral in \cite{Freidel:2005bb,Freidel:2005me}. This aspect of gravity in 3 dimensions will be the central subject of the studies presented in this paper.

In any spacetime dimension Poincar\'e group and (Anti) de Sitter group are symmetries of classical vacuum spacetimes of gravity with zero, (negative) positive cosmological constant.  More than half a century ago it was realized however that there are circumstances that the symmetry of particular configurations of gravitational field is much larger. It was shown that the symmetries of the asymptotically flat gravitational field near null infinity form an infinite dimensional group, called the Bondi, Metzner, Sachs (BMS) group \cite{Bondi:1962px,Sachs:1962wk,Sachs:1962zza}, which contains the Poincar\'e group as its subgroup. A natural question then arises what are the quantum deformations of the BMS group. We addressed this problem in the case of zero cosmological constant in the recent papers \cite{Borowiec:2020ddg,Borowiec:2018rbr}. 

In these papers we found a class of quantum deformations of BMS algebra\footnote{To be more precise, we were dealing with what is sometimes called the extended BMS which includes superrotations in contrast to the original BMS. Some authors also considered a further variant, the "generalized" BMS \cite{Campiglia:2014yka}.}. Technically, we started with the twist deformation of Poincar\'e subalgebras of BMS algebra, extending them to the whole of BMS and obtaining in this way a Hopf--BMS algebra. Here we want to extend this analysis to the case of non-vanishing cosmological constant.

In this paper we consider deformations of BMS algebra of  3-dimensional gravity model of non-zero cosmological constant $\Lambda$-BMS algebra. 

We are interested here in 3 dimensional models because, as mentioned above, it was shown using different approaches that deformed symmetries play an important role in 3 dimensional gravity. In fact, it was established in \cite{Cianfrani:2016ogm} using the non-perturbative methods of Loop Quantum Gravity that the symmetries of the Euclidean Anti de Sitter quantum spacetime in 3 spacetime dimensions is described by a quantum group. More precisely, that paper considers the algebra of the six conserved gravitational charges that form the Euclidean Anti de Sitter algebra. It is then shown by explicit computation that upon quantization these charges become operators, forming the  $SO_q(3,1)$ algebra, with the deformation parameter $q=\exp(i\hbar G\sqrt{|\Lambda|}/2)$. 
In the $\Lambda\rightarrow0$ limit this algebra contracts to  the 3 dimensional (Euclidean) version of $\kappa$-Poincar\'e algebra \cite{ch:II-1Lukierski:1991pn,ch:II-1Lukierski:1992dt,ch:II-1Lukierski:1993wx,ch:II-1Majid:1994cy}. It is believed that an analogous result holds in the case of Lorentzian AdS. This result is a major motivation for investigating deformations BMS  algebra with infinite number of charges in 3-dimensional asymptotically AdS spacetime. Indeed if the Anti de Sitter subalgebra of these charges is getting deformed by non-perturabative quantum gravity effects it can be expected that by the same token the whole algebra of asymptotic charges is deformed as well. It is of interest therefore to list and investigate in some details a class of quantum deformations of the 3-dimensional BMS algebra in the case of negative cosmological constant.

On the more technical side the undertaking to investigate a class of deformations of AdS-BMS algebras is further motivated by the recently obtained complete classification of deformations of (Anti) de Sitter algebras in three dimensions \cite{Borowiec:2017apk} (the discussion of the corresponding contractions $\Lambda\rightarrow 0$ can be found in \cite{Kowalski-Glikman:2019ttm}). Using this classification in the present paper we find a class of deformations of 3-dimensional $\Lambda$-BMS algebra.

Another reason why we choose to investigate here the simpler 3-dimensional $\Lambda-$BMS algebra is that in 4 dimensions the non-vanishing cosmological constant extension of the BMS algebra cannot have a structure of Lie algebra. More precisely, it has been recently shown in \cite{Safari:2019zmc},  by applying cohomological arguments, that there  does not exist a $\Lambda-$BMS$_4$ Lie algebra containing the 4 dimensional (Anti) de Sitter subalgebra that gives the
BMS$_4$ algebra  as a contraction limit $\Lambda\rightarrow0$. 
In fact, such extension of the BMS$_4$ algebra should presumably have the structure of Lie algebroids, with structure functions instead of structure constants \cite{Safari:2019zmc}, \cite{Compere:2019bua}, \cite{Compere:2020lrt}. As a consequence to deform it we would   need a theory of quantized Lie algebroids, which  is much less developed than the theory of quantized enveloping algebras we are dealing with. 

Note that the deformations discussed in this work deform the coalgebra structure. Deformations and (central) extensions of the algebra sector have been addressed for example in \cite{Safari:2019zmc} for the $\mathfrak{B}_4$ and in \cite{Enriquez-Rojo:2021rtv} for diffeomorphisms on the two-sphere.
%In this paper we will consider only the 3-dimensional model of non-zero cosmological constant $\Lambda$-BMS algebra. We are motivated here by the recently obtained complete classification of deformations of (Anti) de Sitter algebras in three dimensions \cite{Borowiec:2017apk} (the discussion of the corresponding contractions $\Lambda\rightarrow0$ can be found in \cite{Kowalski-Glikman:2019ttm}). Using this classification in the present paper we find a class of deformations of 3-dimensional $\Lambda$-BMS algebra\footnote{Another reason why we choose to investigate here the simpler 3-dimensional $\Lambda$-BMS algebra is that in 4 dimensions the non-vanishing cosmological constant extension of the BMS algebra has the structure of a Lie algebroid, with structure functions instead of structure constants \cite{Safari:2019zmc}, \cite{Compere:2019bua}, \cite{Compere:2020lrt}. Despite some attempts to generalize the notions of quantum groups/Lie bialgebras to algebroids \cite{Brzezinski:2002} there is no established concept for deformations of Lie algebroids.}
%\ab{Compere  suggests Lie algebroids. Hopf algebroid is a different concept. }

The plan of the paper is as follows. In the next section we briefly recall the structure of asymptotic symmetries $\Lambda-${BMS}$_3$ algebra in 3 dimensions and interpret the algebras for different signs of cosmological constant as two real forms of a complex algebra. In Section 3 we discuss Lie bialgebras and deformations, first in general terms and then in the specific case of interest of two copies of Witt algebra $\Lambda-${BMS}$_3\simeq\mathfrak{W} \oplus \mathfrak{W}$. Section 4 is devoted to discussion of twist deformations, their classifications and contractions. A schematic overview is presented in Figure 1. We conclude our paper with some remarks on one-sided Witt algebra and specialization in Section 5.

\section{Asymptotic Symmetries of Spacetimes with Cosmological Constant}

In this section we describe the structure of the $\Lambda-{\rm BMS}_3$ algebra of asymptotic symmetries. An extensive discussion of this algebra can be found in \cite{Oblak:2016eij} and \cite{Compere:2018aar}, which contain also references to other works.

\subsection{Asymptotic Symmetries in 3D}\label{sec12}

The study of asymptotic symmetries is usually carried out by starting from a general metric with given asymptotic structure (usually asymptotically Minkowski or (Anti) de Sitter) and imposing fall-off conditions on the expansion coefficients close to the asymptotic boundary. Then one looks for vector fields preserving the form of the asymptotic expansion.
In the three dimensional  asymptotically AdS spacetime such vector fields have the form $\xi_{f, R} = f \partial_u + R \partial_z$ where $R = R(z), f = T(z) + u \partial_z R$ and their algebra reads \cite{Oblak:2016eij, Compere:2018aar}
\begin{align}
[\xi_1, \xi_2] & = \hat \xi \equiv \hat f \partial_u + \hat R \partial_z \\
\hat f & = R_1 \partial_z f_2 + f_1 \partial_z R_2 - (1 \leftrightarrow 2) , \quad \hat R & = R_1 \partial_z R_2 - \Lambda f_1 \partial_z f_2  - (1 \leftrightarrow 2).
\end{align}
Parametrizing $f_m \equiv T_m = z^{m+1}$ and $R_m \equiv l_m = z^{m+1}$ we find    
\begin{align}\label{3dbms}
[l_m, l_n] & = (m-n) l_{m+n}, \quad [l_m, T_n] =  (m-n)T_{m+n}, \\
[T_m, T_n] & = -\Lambda(m-n) l_{m+n}\label{3dalg}
\end{align}
which in the contraction limit $\Lambda \rightarrow 0$ gives the usual BMS$_3$ algebra \eqref{3dbms}. Depending on the sign of $\Lambda$ \eqref{3dbms}-\eqref{3dalg} describes two different real algebras, into which one can embed  the finite 3D (Anti) de Sitter algebra\footnote{AdS corresponds to $\Lambda < 0$ and dS to $\Lambda >0$ and we choose the metric to be $\eta_{+-} = 1, \eta_{22} =-1$.
From the point of view of four-dimensional geometry one can set $\eta_{33}=- \eta_{+-} \Lambda$ and $K_\pm=M_{\pm 3}, K_2=M_{23}$.} generated by the generators $K_2$, $K_\pm$, $M_{+-}$, $M_{\pm2}$ satisfying the algebra
\begin{align}\label{ads3}
[K_+, K_- ] =& -  \eta_{+-} \Lambda M_{+-}, \quad [K_{\pm}, K_2] = -  \eta_{+-} \Lambda M_{\pm 2}, \\
[M_{+2}, M_{-2}] =& -\eta_{22}  M_{+-}, \quad [M_{+-}, M_{\pm 2}] = \pm  \eta_{+-} M_{\pm 2}, \quad [M_{+-}, K_{\pm}] = \pm  \eta_{+-} K_{\pm}, \\
[M_{\pm 2}, K_2] =& \eta_{22} K_{\pm}, \quad [M_{\pm 2}, K_{\mp}] = -  \eta_{+-}  K_2, \label{ads3l}
\end{align}
in an infinitely many distinct ways by identifying (we rescale $\Lambda \rightarrow n^2 \Lambda$)
\begin{align}\label{pt}
 K_2 =  \frac{T_0}{n}, \quad K_{\pm} = - \frac{T_{\pm n}}{\sqrt{2} n}, \quad M_{+-} =  \frac{l_0}{n}, \quad M_{\pm 2} = \mp \frac{1}{\sqrt{2}n} l_{\pm n}
\end{align}
% \ju{remove this part}
% It can also be obtained from the four dimensional Lorentz algebra
% \begin{align}
% [M_{\mu \nu} , M_{\rho \lambda}] = i (\eta_{\mu \lambda} M_{\nu \rho} -  \eta_{\nu \lambda} M_{\mu \rho} + \eta_{\nu \rho} M_{\mu \lambda} - \eta_{\mu \rho} M_{\nu \lambda})
% \end{align}
% if $M_{\pm 3}, M_{2 3}$ is identified with $\tilde K_{\pm },\tilde K_2$. Here the metric is given by $\eta_{+-} = 1, \eta_{22} =-1, \eta_{33} = -1$. $\tilde K_i$ also has to be rescaled 
% \begin{align}
% K_i = R^{-1} \tilde K_i
% \end{align}
% where $R^{-1} = \sqrt{|\Lambda|}$ is the inverse (anti) de Sitter radius and $i = \pm, 2$. Alternatively one could set the metric component $\eta_{33}$ to $\Lambda$.

%Because of the absence of off-diagonal metric elements on the circle the bracket \eqref{3dalg} has the structure of a Lie algebra and not of an Lie algebroid (in a non-trivial sense).\jkg{What does it mean?}
Furthermore, the $\Lambda$-BMS$_3$ algebra \eqref{3dbms}-\eqref{3dalg} is isomorphic to two copies of the Witt algebra $\mathfrak{W} \oplus \mathfrak{W}$ via
\begin{align}\label{cwiso}
L_m & = \frac{1}{2}\left( l_m+ \frac{1}{\sqrt{-\Lambda}} T_m\right), \quad \bar L_m = \frac{1}{2}\left( l_m- \frac{1}{\sqrt{-\Lambda}} T_m\right) \\
\Rightarrow [L_m, L_n] & = (m-n) L_{m+n}, \quad [\bar L_m, \bar L_n] = (m-n) \bar L_{m+n}, \quad [L_m, \bar L_n] = 0. \label{cwalg}
\end{align}
The isomorphism \eqref{cwiso} is  complex for positive $\Lambda$ and therefore \eqref{cwalg} has to be seen as a complex algebra with different real forms (cf. next section).
As before, it is also easy to see from  \eqref{cwiso}  that there are infinitely many embeddings of the (A)dS algebra into the $\Lambda$-BMS$_3$, i.e. one shows that the $\mathfrak{o}(4, \mathbb{C})\equiv \mathfrak{sl}(2, \mathbb{C})\oplus \mathfrak{sl}(2, \mathbb{C})$ is { multiply} embedded in the two copies of the Witt algebra via
\begin{align} \label{embed1}
L_0,  L_{\pm n}, \bar L_0, \bar L_{\pm n}, \\
L_n \rightarrow \frac{L_n}{n}, \quad \bar L_n \rightarrow \frac{\bar L_n}{n},\quad n=1,2,\ldots\,. \label{embed2}
\end{align} 
Using the isomorphism \eqref{cwiso} this translates to {the family} of embeddings
\begin{align}
l_0, l_{\pm n}, T_0, T_{\pm n}
\end{align}
with the rescaling 
\begin{align}\label{rescale}
l_n \rightarrow \frac{l_n}{n}, \quad \sqrt{-\Lambda} \rightarrow n \sqrt{- \Lambda}, \quad n=1,2,\ldots\,.
\end{align}
Alternatively one can rescale $\eta_{\mu \nu}$ instead of the generators $l_m$, i.e. instead of \eqref{pt} one would have the same relations with $n = 1$ and $\Lambda$ is not rescaled.

\subsection{Real Forms}

A real Lie algebra is naturally defined as a real vector space with Lie bracket determined by real structure constants. However, for the purpose of quantum deformation one needs another, equivalent definition, which is based on the notion of a real form of a complex Lie algebra (see e.g. \cite{Borowiec:2017apk} and references therein). Thus real form is a pair $(\mathfrak{g, \dagger})$ where $\mathfrak{g}$ is a complex Lie algebra and $\dagger:\mathfrak{g}\rightarrow\mathfrak{g}$ denotes antilinear involutive antiautomorphism mimicking  Hermitian conjugation, see below. If the structure constants are real then the natural choice is $X_a^\dagger=-X_a, X_a\in gen(\mathfrak{g})$. 
Favorite physicist convention is to establish imaginary structure constants and Hermitian generators $Y_a^\dagger=Y_a$, where $Y_a=i X_a$.

For example, the simple  $\mathfrak{sl}(2,\mathbb{C})$ Lie algebra admits (up to an isomorphism) two real forms:
noncompact $\mathfrak{sl}(2,\mathbb{R})\sim \mathfrak{o}(1,2)\sim \mathfrak{su}(1,1)$ and compact $\mathfrak{o}(3)\sim \mathfrak{su}(2)$
\footnote{Different notational coventions  reference to different $\star$ realisations or different system of generators.}.
Accordingly, the semisimple $\mathfrak{o}(4,\mathbb{C})=\mathfrak{sl}(2,\mathbb{C})\oplus \mathfrak{sl}(2,\mathbb{C})$ admits four non-isomorphic real forms: Euclidean, Lorentzian, Kleinian and quaternionic  \cite{Borowiec:2015nlw,2015nlw2}. Each of them can be extended to the real form of the infinite-dimensional  $\Lambda$-BMS  algebra. However, in  view of possible physical applications we are interested here in Lorentzian and Kleinian type. They correspond to de Sitter and anti de Sitter algebras of 3-dimensional Lorentzian spacetime $\mathbb{R}^{1,2}$.

Note that while \eqref{ads3}-\eqref{ads3l} describes two different real  algebras with $\Lambda \lessgtr 0$, in \eqref{cwalg} there is only one complex algebra with two different reality conditions.
If we consider the subalgebra $\mathfrak{o}(4, \mathbb{C})$ spanned by $L_0, L_{\pm 1}, \bar L_0, \bar L_{\pm 1}$
\begin{align}\label{o4}
[L_0, L_{\pm 1}] = \mp L_{\pm 1}, \quad [L_{+1}, L_{-1}] = 2 L_0, \\
[\bar L_0, \bar L_{\pm 1}] = \mp \bar L_{\pm 1}, \quad [\bar L_{+1}, \bar L_{-1}] = 2 \bar L_0, 
\end{align}
it is related to the standard Cartan-Weyl form
\begin{align}\label{o4cw}
[H, E_{\pm }] = \pm E_{\pm }, \quad [E_{+}, E_{-}] = 2 H, \\
[\bar H, \bar E_{\pm }] = \pm \bar E_{\pm }, \quad [\bar E_{+}, \bar E_{-}] = 2 \bar H, 
\end{align}
 via\footnote{It is worth noticing that Cartan-Weyl generators of $\mathfrak{sl}(2, \mathbb{R})$ can be also considered as  light-cone generators of $\mathfrak{o}(1,2)$ through the identification: $M_{+-}=H, M_{1\pm}=E_\pm$ with non-diagonal metric components $\eta_{+-}=\eta_{-+}=1$ and diagonal one $\eta_{22}=-2$ (cf. \eqref{ads3}).}
% \ab{why not the most natural identification without explicit  complex coeff\\
% 	 $H= -L_0, \quad E_{\pm} = \pm\lambda^{\pm 1}\,L_{\pm 1}, \lambda\neq 0$}
\begin{align}
H= -L_0, \quad E_{\pm} = i L_{\pm 1}, \quad \bar H = - \bar L_0, \quad \bar E_{\pm} = i \bar L_{\pm 1}.
\end{align}

% \ab{why not the most natural identification without complex coeff.}
%\footnote{It is worth noticing that Cartan-Weyl generators of $\mathfrak{sl}(2, \mathbb{R})$ can be also considered as  light-cone generators of $\mathfrak{o}(1,2)$ through the identification: $M_{+-}=H, M_{1\pm}=E_\pm$ with non-diagonal metric components $\eta_{+-}=\eta_{-+}=1$ and diagonal one $\eta_{22}=-2$ (cf. \eqref{ads3}).}
%\begin{align}
%H= -L_0, \quad E_{\pm} = i L_{\pm 1}, \quad \bar H = - \bar L_0, \quad \bar %E_{\pm} = i \bar L_{\pm 1}.
%\end{align}
In \eqref{o4} there are two real forms that correspond to the AdS and dS case resprectively. For negative $\Lambda$, i.e. the AdS case we have from \eqref{cwiso} that 
\begin{align}
L_m^{\dagger} & = - L_m, \quad \bar L_m^{\dagger} = -\bar L_m \\
\Leftrightarrow H^{\dagger} = - H, \quad E_{\pm}^{\dagger} = E_{\pm}, \quad \bar H^{\dagger} & = - \bar H, \quad \bar E_{\pm}^{\dagger} = \bar E_{\pm}
\end{align}
and restrained to the subalgebra this defines two copies of the real form $\mathfrak{sl}(2, \mathbb{R}) \simeq \mathfrak{o}(2,1)$. Thus this real form corresponds to the Kleinian algebra $\mathfrak{o}(2,2) \simeq \mathfrak{o}(2,1) \oplus \bar {\mathfrak{o}}(2,1)$.

The other case with positive $\Lambda$ yields
\begin{align}
L_m^{\ddagger} & = -\bar L_m, \quad \bar L_m^{\ddagger} = -L_m, \\
\Leftrightarrow H^{\ddagger} & = - \bar H, \quad E^{\ddagger}_{\pm} = \bar E_{\pm},
\end{align}
i.e. the Lorentzian real form when restricted to the $\mathfrak{o}(4, \mathbb{C})$ subalgebra. 
This can be identified with the real structures listed in \cite{Borowiec:2017apk} in the last line of eq.(4.13) and eq.(4.14) with the automorphism $E_{\pm} \rightarrow - E_{\pm}, \bar E_{\pm} \rightarrow - \bar E_{\pm}$.  Note that this automorphism of the $\mathfrak{sl}(2, \mathbb{C})$ 
\begin{align}\label{auto}
\Phi (E_{\pm}) = - E_{\pm}, \quad \Phi(H) = H, \rightarrow \Phi (L_{\pm 1}) = - L_{\pm 1}, \quad \Phi(L_0) = L_0,
\end{align}
can be extended uniquely to an automorphism of the Witt algebra.

As mentioned above for the algebras \eqref{ads3}-\eqref{ads3l} we have only one reality condition
\begin{align}
K_{\mu}^{\dagger} = K_{\mu}, \quad M_{\mu \nu}^{\dagger} = M_{\mu \nu}.
\end{align}

\subsection{Algebra of Surface Charges}\label{secsc}

By the Noether theorem the new-found symmetries correspond to conserved quantities, although this relation is more involved in the case of gauge symmetries (see \cite{Oblak:2016eij}). For example in the flat case one obtains infinitely many charges parametrized by supertranslations $f(z, \bar z)$ \cite{Strominger:2017zoo}
\begin{align}
    Q[f] = \frac{1}{4 \pi G} \int d^2 z \gamma_{z \bar z} f m_B,
\end{align}
where the integral is defined over a boundary of a spacelike slice and $m_B$ is the Bondi mass aspect. These charges generate the symmetry transformations and are thus tightly linked to the algebra derived above. In fact the Poisson bracket of the charges has to coincide with the Lie bracket of the algebra generators up to a constant, i.e. it is in general a central extension of the algebra \cite{Oblak:2016eij}. 
For the Witt algebra it is well known that the central extension is uniquely given by the Virasoro algebra ($\mathfrak{Vir}$) satisfying the commutation relations
\begin{align}\label{centr}
    [L_m, L_n] = (m-n) L_{m+n} + \frac{c_L}{12} (m^3-m)\delta_{m+n, 0}.
\end{align}
Furthermore, two copies of the Virasoro algebra consitute the only possibility for a central extension of $\mathfrak{W} \oplus \mathfrak{W}$ and Brown and Henneaux calculated in their seminal paper \cite{Brown:1986nw} that 
\begin{align}
    c_L = c_{\bar L} = \frac{3}{2 G \sqrt{- \Lambda}}.
\end{align}
{It should be noticed that only the first embedding $(L_0, L_{\pm 1})$ of $\mathfrak{sl}(2)$ algebra is not affected by the presence of central charge in  }\eqref{centr}. The higher order embeddings, e.g. $(L_0, L_{\pm 2})$, have to take into account the central charge.

\section{Lie Bialgebras and Deformation}

Recall that a Lie bialgebra is a Lie algebra $\mathfrak{g}$ with a cobracket $\delta : \mathfrak{g} \rightarrow \mathfrak{g} \otimes \mathfrak{g}$ satisfying the cocycle condition \cite{chari1995guide} 
\begin{align}
\delta([x, y]) = [ x \otimes 1 + 1 \otimes x, \delta(y)] - [ y \otimes 1 + 1 \otimes y, \delta(x)] 
\end{align}
and the dual version of the Jacobi identity, the so-called co-Jacobi identity
\begin{align}\label{cojac}
\text{Cycl}((\delta \otimes \text{id} )) \delta (x) = 0
\end{align}
with $\text{Cycl}(a \otimes b \otimes c) = a \otimes b \otimes c + c \otimes a \otimes b + b \otimes c \otimes a$.

 A coboundary Lie bialgebra has a cobracket defined by a classical r-matrix $r \in\bigwedge^2 \mathfrak{g}$ via
\begin{align}
\delta_r (x) = [x \otimes 1 + 1 \otimes x, r]
\end{align}
and $\delta_r$ satisfies the co-Jacobi identity iff $r = a \wedge b$ fulfills the modified classical Yang-Baxter equation
\begin{align}\label{mcybe}
[[r, r]] \equiv [r_{12}, r_{13}] + [r_{12}, r_{23}] + [ r_{13}, r_{23}]  = \Omega
\end{align}
where $r_{12} = a \otimes b \otimes 1 - b \otimes a \otimes 1$ and $\Omega$ has to be ad-invariant in $\mathfrak{g}$. If the rhs of \eqref{mcybe} vanishes the Lie bialgebra is called triangular.

A $*$-Lie bialgebra over a real form of a complex algebra with an involution $*$ is a Lie bialgebra that is a $*$ vector space and bracket and cobracket are $*$ homomorphisms. The latter condition implies for coboundary Lie bialgebras defined by an r-matrix $r$ that
\begin{align}\label{invoc}
r^{* \otimes *} = -r,
\end{align}
where $(a \otimes b)^{* \otimes *} := a^* \otimes b^*$.

We recall that two $r$-matrices $r_1, r_2\in \mathfrak{g}\wedge \mathfrak{g}\subset \mathfrak{g}\otimes \mathfrak{g}$ are called equivalent if there exists a Lie algebra automorphism $\phi\in Aut (\mathfrak{g})$ such that
$(\phi\otimes \phi)\, (r_1)=r_2$. Equivalent $r$-matrices provide isomorphic Lie bialgebra structures on $\mathfrak{g}$.  Choosing  Lie subalgebras $\mathfrak{h}\subset \mathfrak{g}$ and 
$r_1, r_2\in \mathfrak{h}\wedge \mathfrak{h}\subset \mathfrak{g}\wedge \mathfrak{g}$ one can ask  now whether 
$\mathfrak{h}$-equivalence implies $\mathfrak{g}$-equivalence. The answer is
not obvious since in general an automorphism of $\mathfrak{h}$ does not extend to the automorphism of the full algebra $\mathfrak{g}$. Therefore, the classification problem
 depends on the choice of an algebra we are interested in, instead of just the minimal subalgebra generated by the $r$-matrix itself\footnote{This problem involves only triangular case. Non-triangular $r$-matrices
can not be promoted from subalgebras, c.f. \eqref{mcybe}.}.   Similarly, if $(\mathfrak{g},\star)$ is a real form of a complex Lie algebra $\mathfrak{g}$ then $Aut(\mathfrak{g},\star)\subset Aut(\mathfrak{g})$. Therefore, equivalent complex $r$-matrices may not be equivalent as real ones.

Lie bialgebras can be considered as infinitesimal versions of Hopf algebras, i.e. unitary algebras with a compatible coproduct $\Delta$, counit $\varepsilon$ and an antipode $S$ generalizing the inverse (cf.\ \cite{chari1995guide, klimyk2011quantum} for an extensive treatment of Hopf algebras/quantum groups). In particular the cobracket is related to the coproduct via
\begin{align}
\delta = \underset{\kappa \rightarrow \infty}{\text{lim}} \frac{\Delta - \Delta^{\text{op}}}{1/\kappa},
\end{align}
where $1/\kappa$ is the deformation parameter (see below).
Starting from any Lie algebra $\mathfrak{g}$ one can generically construct a Hopf algebra $H$ by considering the universal enveloping algebra $U\mathfrak{g}$ with 
\begin{align}
\Delta_0(x) = x \otimes 1 + 1 \otimes x, \quad \varepsilon (x) = 0, \quad S_0(x) = -x .
\end{align}
Non-trivial coalgebra structures can be obtained by a deformation, i.e. $U \mathfrak{g}$ is first topologically extended to $U \mathfrak{g} [[1/\kappa]]$ with the so-called $h$-adic topology to include formal power series in the deformation parameter $1/\kappa$. 
If $\mathfrak{g}$ admits a triangular Lie bialgebra structure such a deformation can be obtained by a twisting procedure, i.e. then a twist $\mathcal{F} \in H \otimes H$ satisfying the 2-cocycle condition
\begin{align}\label{coc-cond}
\mathcal{F}_{12} ( \Delta_0 \otimes 1) (\mathcal{F}) = \mathcal{F}_{23} (1 \otimes \Delta_0)(\mathcal{F})
\end{align}
exists and defines a deformed coproduct via
\begin{align}
\Delta_{\mathcal{F}} = \mathcal{F} \Delta_0 \mathcal{F}^{-1}.
\end{align}
In the following we will find that all possible deformations are of this form in the $\mathfrak{W}\oplus \mathfrak{W}$ algebra with the help of Lie algebra cohomology.

\subsection{Cohomology}
It is well known that the relation
\begin{align}\label{coco}
\delta ([L_m, L_n]) = [L_m \otimes 1 + 1 \otimes L_m, \delta(L_n)] - [L_n \otimes 1 + 1 \otimes L_n, \delta(L_m)] 
\end{align}
is the condition that $\delta$ is a 1-cocycle of the Chevalley-Eilenberg cohomology \cite{etingof2010lectures}. Recall that this cohomology is constructed on the vector spaces of cochains $C^n = \text{Hom} (\Lambda^n \mathfrak{g}, V)$ where $V$ is a module of the Lie algebra $\mathfrak{g}$ (in our case $V = \bigwedge^2 \mathfrak{g}$). The coboundary operators $\partial_n : C^n \rightarrow C^{n+1}$ are given by
\begin{align}
\partial_n (f)(x_1 \wedge ... \wedge x_n+1) = & \sum_{i=1}^{n+1} (-1)^i x_i \triangleright f (x_1 \wedge ... \wedge \hat {x_i} ... \wedge x_{n+1}) \nonumber \\
&+ \sum_{i<j} (-1)^{i+j} f( [x_i, x_j] \wedge x_1 ... \wedge \hat{x_i} ... \wedge \hat{x_j} ... \wedge x_{n+1})
\end{align}
where $\hat{x_i}$ means that the $i$-th tensor leg is dropped and $\triangleright$ denotes the right action on the module. We denote the cohomology groups by $\text{Ker } \partial_n / \text{Im } \partial_{n-1} \equiv H^n(\mathfrak{g}, V)$.

If the first cohomology $H^1(\mathfrak{W} \oplus \mathfrak{W}, \bigwedge^2( \mathfrak{W} \oplus \mathfrak{W}))$ vanishes it would follow that all Lie bialgebras in $\mathfrak{W} \oplus \mathfrak{W}$ are coboundary. The result is even a bit stronger as not all cocycles of the cohomology define Lie bialgebras but only those which additionally fulfill the co-Jacobi identity. 

In the theory of finite dimensional Lie algebras a fundamental result is the Whitehead lemma which states that all cohomology groups $H^n(\mathfrak{g}, V)$ for finite-dimensional semi-simple $\mathfrak{g}$ and $V$ vanish. However, since $\mathfrak{W}$ is not finite dimensional the lemma is not applicable here. Thus we prove the following theorem in Appendix \ref{appA}
\begin{theorem}
The first cohomology $H^1(\mathfrak{W} , \bigwedge^2 \mathfrak{W} )$ of the Witt algebra with values in the exterior product of the adjoint module is zero.
\end{theorem}

Using this theorem one can also prove the following
\begin{theorem}
The first cohomology $H^1( \mathfrak{W} \oplus \mathfrak{W}, \bigwedge^2 (\mathfrak{W} \oplus \mathfrak{W}))$ is zero.
\end{theorem}

In \cite{Junbo:2010} the authors independently prove a slightly more general result than our first theorem at the cost of a longer proof. The proofs presented here conceptually follow the proof of $H^1( \mathfrak{W}, \mathfrak{W}) = \{0\}$ in \cite{Ecker:2019thw}.

As stated above the second theorem establishes that all Lie bialgebra structures are coboundary and we now just need to show that the corresponding deformations are all given by a twist.

\section{Twist Deformation and Classification}

In the recent paper \cite{Borowiec:2020ddg} several twists were considered for the BMS algebra in three and four dimensions (denoted by $\mathfrak{B}_{3}$ and $\mathfrak{B}_{4}$, respectively). It was noticed that all deformations from coboundary Lie bialgebras have to be triangular since there is no ad-invariant element in $\wedge{}^3 \mathfrak{B}_{3}$ and $\wedge{}^3\mathfrak{B}_{4}$. The same observation also holds for $\mathfrak{W} \oplus \mathfrak{W}$. Since in four dimensions the $\Lambda$-BMS is a Lie algebroid it is not known how a suitable concept of quantum group can be defined on it. In three dimensions, however, we can investigate the possible twists in a similar way. 

As motivated earlier we will focus on $\Lambda < 0$ in the following, if not stated otherwise.
In the contraction limit one can identify the generators $K_i$ with the momenta of the $\mathfrak{B}_3$, i.e. 
\begin{align}\label{contr}
 \underset{\Lambda \rightarrow 0}{\text{lim}}K_i= P_i . 
\end{align}

Let us first consider  the three dimensional Poincar\'e ($\mathfrak{P}_3$).
The abelian twist and the Jordanian twist, corresponding to the r-matrices
\begin{align}\label{rm}
r_J = i \eta^{+-} M_{+-} \wedge P_+, \quad r_A = -i \eta^{+-} M_{+-} \wedge P_2
\end{align}
are then also viable if $P_i$ is replaced with $K_i$, i.e. the r-matrices are triangular and the twists satisfy the 2-cocycle condition.
Also the r-matrix associated with the light-cone $\kappa$-Poincar\'e
\begin{align}\label{lcr}
r_{\text{LC}} = iM_{+-} \wedge K_+ - i M_{+2} \wedge K_2,
\end{align}
is triangular in three dimensions and when expressed in terms of $L_m, \bar L_m$
\begin{align}\label{lcr2}
r_{\text{LC}} = - \frac{i \Lambda \sqrt{2}}{n} (L_0 \wedge L_n - \bar L_0 \wedge \bar L_n)
\end{align}
and it is apparent that it coincides with $r_{II} (\zeta =0)$ from \cite{Borowiec:2017apk} and \cite{Kowalski-Glikman:2019ttm} where all classes of available twists of $\mathfrak{o}(4)$ were obtained.
Demanding triangularity we are left with the following r-matrices from the classification
\begin{align}\label{class1}
r_{I} & = \chi (E_+ - \bar E_+) \wedge (H + \bar H), \\
r_{II} & = \chi E_+ \wedge H + \bar \chi \bar E_+ \wedge \bar H + \zeta E_+ \wedge \bar E_+, \\
r_{III} & = \eta H \wedge \bar H. \\
r_{V} & = \bar \chi \bar E_+ \wedge \bar H + \rho H \wedge \bar E_+.\label{class2}
\end{align}
The abelian twist corresponds to $r_{III}$ and the Jordanian twist to $r_{I}$. 

In general there are also other classical r-matrices in $\mathfrak{W} \oplus \mathfrak{W}$ and the full classification is not known even for the Witt algebra \cite{Ng:2000}. For example it is easy to see that r-matrices of the form
\begin{align}
r = \left( \sum_i \alpha_i L_i \right)\wedge \bar L_m    
\end{align}
are triangular.

However, one has to take into account that the asymptotic symmetry is spontaneously broken in the bulk in the sense that the vacua related by supertranslations and superrotations are physically distinguishable \cite{Strominger:2017zoo}. 
There is a correspondence between these vacua and the embeddings of Poincar\'e subalgebras which leave the associated vacuum invariant.
Therefore we require that the restriction of the Hopf algebra deformed by the twist associated with a given r-matrix to an embedding is a sub Hopf algebra and we are interested in r-matrices of the form \eqref{class1}-\eqref{class2} where $\{H, E_{\pm}, \bar H, \bar E_{\pm}\}$ is replaced with the embedding.
Note that while in the case of positive $\Lambda$ the involution mixes left and right-handed elements this is not happening for negative $\Lambda$, leaving the potential possibility to use different embeddings for them.

The classification \eqref{class1}-\eqref{class2} is defined up to automorphisms of the $\mathfrak{o}(4)$ but there might be inequivalent r-matrices that are related by $\text{Aut}(\mathfrak{o}(4))$ that do not extend to $\text{Aut}(\mathfrak{W} \oplus \mathfrak{W})$.

Therefore, in Appendix \ref{app-b} the classification of triangular r-matrices on $\mathfrak{o}(4)$ is revised  along the lines of \cite{Borowiec:2015nlw,2015nlw2} but using only the $\mathfrak{W} \oplus \mathfrak{W}$ 
automorphisms ($\text{Aut}(\mathfrak{W} \oplus \mathfrak{W})$)
\begin{align}\label{au1}
\varphi_{(\gamma,\bar\gamma,\epsilon,\bar\epsilon)}(L_m) = \gamma^m \epsilon L_{\epsilon m}, \quad \varphi_{(\gamma,\bar\gamma,\epsilon,\bar\epsilon)}(\bar L_m) = \bar \gamma^m \bar \epsilon  \bar L_{\bar \epsilon m}, \\
\varphi'(L_m) = \bar L_m , \quad \varphi'(\bar L_m) = L_m, \label{au2}
\end{align}
where $0\neq \gamma,\bar\gamma\in \mathbb{C}, \epsilon,\bar\epsilon=\pm 1$.
As a result we obtain the following classes of r-matrices
\begin{align}\label{classa}
r_{1'} & = \beta (L_1 + L_{-1} + 2 L_0) \wedge ( \bar L_1 + \bar L_{-1} + 2 \bar L_0) + a_1 + \bar a_1, \\
r_{2'} & = \beta L_{1} \wedge \bar L_0 + a_2 , \\
r_{3'} & = \beta (L_1 + \epsilon L_0 + \epsilon' L_{-1} ) \wedge (\bar L_1  + \bar L_{-1} + 2 \bar L_0) + \bar a_1 + (1- \epsilon) (1- \epsilon') a_2, \\
r_{4'} & = \beta_1  L_1 \wedge \bar L_1 + \beta_2  (L_1+ \bar L_1) \wedge (L_0 +  \bar L_0 ) , \\
r_{5'} & = \beta (L_1 + L_{-1}) \wedge ( \bar L_1 + \bar L_0) + a_1, \\
r_{6'} & = (\beta L_1 + \beta_0 L_0  + \epsilon \beta L_{-1}) \wedge ( \bar \beta \bar L_1 + \bar \beta_0 \bar L_0 + \bar \epsilon \bar \beta \bar L_{-1}), \\
r_{7'} & = L_1 \wedge ( \bar \beta \bar L_1 + \bar \beta \bar L_0 + \epsilon \bar \beta \bar L_{-1}) + a_2, \\
r_{8'} & = \beta L_1 \wedge \bar L_{1} + a_2 + \bar a_2, \label{classb}
% r_{9'} & = \alpha_1 L_1 \wedge \bar L_0 + \alpha_2 L_0 \wedge \bar L_1 + \alpha_1 L_1 \wedge L_0 + \alpha_2 \bar L_1 \wedge \bar L_0,
\end{align}
where $\epsilon, \epsilon', \bar \epsilon \in \{0,1\}$ and
\begin{align}
a_1 & = \alpha ( L_1 \wedge L_0 +  L_1 \wedge L_{-1} -L_{-1} \wedge L_0), \quad 
a_2  = \alpha L_1 \wedge L_0, \\
\bar a_1 & = \bar \alpha (\bar L_1 \wedge \bar L_0 + \bar L_1 \wedge \bar L_{-1} - \bar L_{-1} \wedge \bar L_0), \quad 
\bar a_2  =\bar \alpha \bar L_1 \wedge \bar L_0.
\end{align}

In the case of complex $\mathfrak{W} \oplus \mathfrak{W}$ all the parameters in \ref{classa}-\ref{classb} can take values in $\mathbb{C}$ but for the real forms associated with the involutions $\dagger, \ddagger$ the condition \eqref{invoc} constrains the choice of parameters. For $\dagger$ in the classes $r_{1'}$ to $r_{5'}$ and $r_{7'}, r_{8'}$ this enforces $\beta, \beta_1, \beta_2, \alpha, \bar \alpha \in i \mathbb{R}$ and in $r_{6'}$ one can restrict $\beta, \beta_0 \in i \mathbb{R}$ and $\bar \beta, \bar \beta_0 \in \mathbb{R}$ without loss of generality.

For positive $\Lambda$, i.e. the involution $\ddagger$, the reality condition is more restrictive. In particular
\begin{align*}
r_{1'}:& \quad \alpha = \bar \alpha \in i \mathbb{R}, \beta \in \mathbb{R}, \quad r_{2'}: \quad \beta = 0, \alpha = \bar \alpha \in i \mathbb{R}, \\
r_{3'}:& \quad \bar \alpha =0, \epsilon = \epsilon' = 1, \beta \in \mathbb{R}, \quad r_{4'}: \quad \beta_1 \in \mathbb{R}, \beta_2 \in i \mathbb{R}, \\
r_{5'}:& \quad \text{excluded}, \quad r_{6'}: \quad \epsilon = \bar \epsilon, \beta_0 \bar \beta = - \frac{\beta^*}{\beta_0^*} \text{ or } (\beta_0 = \bar \beta_0 =0, \beta, \bar \beta \in \mathbb{R}), \\
r_{7'}:& \quad \text{excluded}, \quad r_{8'}: \quad \beta \in \mathbb{R}, \alpha = \bar \alpha \in i \mathbb{R}.
\end{align*}

As the r-matrices from \eqref{classa}-\eqref{classb} that are not included in \eqref{class1}-\eqref{class2} are at least in  $\mathfrak{o}(4)$ automorphism orbits containing them one can use the inverse of the automorphisms to obtain the full twists. For example $r= (L_1 - L_{-1}) \wedge (\bar L_1  - \bar L_{-1})$ (which is automorphic to $r_{6'}$ with $\beta_0 = \bar \beta_0 =0, \epsilon = \bar \epsilon =1$ ) is automorphic to $L_0 \wedge \bar L_0 \hat{=} r_{III}$ by 
\begin{align}
\varphi(L_1) = -\frac{1}{2}(L_1 + L_{-1}) + L_0, \quad \varphi(L_{-1}) =  -\frac{1}{2}(L_{1} + L_{-1}) -L_0, \quad \varphi(L_0) =  \frac{1}{2} (L_1 - L_{-1}). 
\end{align}
From the abelian twist for $r_{III}$ 
\begin{align}
\mathcal{F}_{III} = \exp( \eta L_0 \wedge \bar L_0)
\end{align}
one then gets the twist
\begin{align}
\mathcal{F} = \left(\varphi^{-1} \otimes \varphi^{-1}\right) \mathcal{F}_{III} = \exp (\eta (L_1 - L_{-1}) \wedge (\bar L_1 - \bar L_{-1})).
\end{align}

\subsection{Twisting of the Coalgebra Sector}

In this section we will explicitly construct the Hopf algebras from an abelian and a Jordanian twist. The abelian twist here has the peculiarity that it consists only of elements that are contained in all embeddings so it does not single out any specific. The Jordanian twist can already be constructed in a very basic example namely the only non-abelian two dimensional algebra
\begin{align}\label{2d}
[X,Y] = Y,
\end{align}
where $\mathcal{F}= \exp (X \otimes \log (1+Y))$ satisfies the 2-cocycle condition \eqref{coc-cond}. Because of the semi-simplicity of the relevant algebras here there is always a Cartan element that diagonalizes the adjoint action and thus a subalgebra of the form \eqref{2d}. Indeed many of the possible twists are of this form or have it as a building block, making it an ideal example to study.

\subsubsection{Abelian Twist}\label{sec411}
The abelian twist can be expressed as
\begin{align}
\mathcal{F}_A = \exp \left(-\frac{i}{\kappa n^2} \Lambda \bar L_0 \wedge L_0 \right) \exp \left(-\frac{i}{\kappa n^2} \Lambda (L_0 \otimes L_0 - \bar L_0 \otimes \bar L_0 )\right)
\end{align}
which factorizes into the twist $\mathcal{F}_{3''}$ from \cite{Borowiec:2017apk} and a factor that only produces symmetric deformations of the coproduct. Explicitly
\begin{align}
\Delta_{\mathcal{F}_{3''}}(L_m) = e^{i\frac{m}{n^2}\Lambda  \bar L_0} \otimes L_m + L_m \otimes e^{-i \frac{m}{n^2}\Lambda  \bar L_0}  \\
\Delta_{\mathcal{F}_{3''}}(\bar L_m) = e^{-i\frac{m}{n^2}\Lambda   L_0} \otimes \bar L_m + \bar L_m \otimes e^{i \frac{m}{n^2}\Lambda L_0} 
\end{align}
and 
\begin{align}\label{copab1}
\Delta_{\mathcal{F}_A}(L_m) = & e^{\frac{i}{\kappa} \frac{m}{n^2}\Lambda (\bar L_0 + L_0)} \otimes L_m + L_m \otimes e^{-\frac{i}{\kappa} \frac{m}{n^2}\Lambda (\bar L_0 - L_0)} \\
\Delta_{\mathcal{F}_A}(\bar L_m) = & e^{-\frac{i}{\kappa} \frac{m}{n^2}\Lambda (\bar L_0 + L_0)} \otimes \bar L_m + \bar L_m \otimes e^{\frac{i}{\kappa} \frac{m}{n^2}\Lambda ( L_0 -  \bar L_0)} \\
\Delta_{\mathcal{F}_A}(L_0) = & L_0 \otimes 1 + 1 \otimes L_0, \quad \Delta_{\mathcal{F}_A}(\bar L_0) = \bar L_0 \otimes 1 + 1 \otimes \bar L_0. \label{copab2}
\end{align}
The antipodes can also be inferred easily from 
\begin{align}\label{antip}
m \circ (S \otimes \text{id}) \circ \Delta = 1 \circ \epsilon 
\end{align}
and turn out to be
\begin{align}
S_{\mathcal{F}_{3''}}(L_m) = &-L_m, \quad S_{\mathcal{F}_{3''}}(\bar L_m) = - \bar L_m, \\ 
S_{\mathcal{F}_A}(L_m) = & -L_m  e^{\frac{2i}{\kappa} \frac{m}{n^2}\Lambda L_0}, \quad S_{\mathcal{F}_A}(\bar L_m) =-\bar L_m  e^{-\frac{2i}{\kappa} \frac{m}{n^2}\Lambda  \bar L_0} 
\end{align}

\subsubsection{Jordanian Twist}
Considering the Jordanian twist
\begin{align}\label{jtwi}
\mathcal{F}_{J, n} = \exp \left( - \frac{1}{n} (L_0 + \bar L_0) \otimes \log \left(1 - \frac{i  \Lambda}{\kappa \sqrt{2}} (L_n - \bar L_n) \right) \right) 
\end{align}
one finds
\begin{align}
\Delta_{\mathcal{F}_{J, n}}(L_0) = & L_0 \otimes 1 +1 \otimes L_0  - \frac{1}{n}(L_0 + \bar L_0) \otimes \frac{d \sigma_n}{d(L_n - \bar L_n)} n L_n \nonumber \\
= & L_0 \otimes 1 + 1 \otimes L_0 - \tilde a (L_0 + \bar L_0) \otimes L_n \Pi_{+n}^{-1},  \\
\Delta_{\mathcal{F}_{J, n}}(\bar L_0) = & \bar L_0 \otimes 1 + 1 \otimes \bar L_0 + \tilde a (L_0 + \bar L_0) \otimes \bar L_n \Pi_{+n}^{-1}, 
\end{align}
where 
\begin{align}
\sigma_n \equiv \log \left(1 + \tilde a (L_n - \bar L_n)\right), \quad \Pi_{+n} = e^{\sigma_n}, \quad \tilde a \equiv \frac{-i \Lambda}{\kappa \sqrt{2}} \label{defpi}
\end{align}
Using \eqref{antip} and the previous equations we find
\begin{align}
S_{\mathcal{F}_{J, n}}(L_0 ) =& - (L_0 + \tilde a (\bar L_0 L_n + L_0 L_n))\frac{\Pi_{+n}^{-1}}{1-  \Pi_{+n}^{-1} (L_n - \bar L_n)} \nonumber \\
=& - (L_0 + \tilde a (\bar L_0 L_n + L_0  L_n)),  \label{SL0} \\
S_{\mathcal{F}_{J, n}}(\bar L_0) =&  - ( \bar L_0 -  \tilde a (\bar L_0 \bar L_n + L_0 \bar L_n)).
\end{align}

For general generators we find 
\begin{align}
\Delta_{\mathcal{F}_{J, n}}(L_m) & = \mathcal{F}_{J, n} (L_m \otimes 1 + 1 \otimes L_m) \mathcal{F}^{-1}_{J, n} \nonumber \\
& = L_m \otimes \Pi_{+ n}^{\frac{m}{n}} +  \sum_{l=0}^{\infty} \frac{1}{l !} \left(\frac{L_0 + \bar L_0}{n } \right)^{l}  \otimes \big[ \sigma_n, \big[..., \big[\sigma_n, L_m\big]...\big] \label{delgen}
\end{align}
with
\begin{align}\label{comm}
[\sigma_n, L_m] =& \bigg[ - \sum_{j = 0}^{\infty} \frac{(- \tilde a)^j}{j} (L_n - \bar L_n)^j, L_m \bigg] \\
\big[\left(L_n- \bar L_n \right)^{j}, L_m \big] =& \sum_{s=0}^{j-1} (L_n - \bar L_n)^s (n-m) L_{m+n} (L_n - \bar L_n)^{j-s} \nonumber \\
 =&\sum_{k=1}^{j} \bigg( \sum_{s_1 = k-1}^{j-1} ... \sum_{s_k = 0}^{s_{k-1}-1} L_{m+ kn} (L_n - \bar L_n)^{j-k} \prod_{p=0}^{k-1} (n-(m+pn)) \bigg),
\end{align}
where in the last line we iteratively commuted the $s$ terms in front of $L_{m+kn}$ to the right. 

From this formula it can be seen that in general there is an infinite number of terms in the coproduct involving a tower of infinitely many different generators. However, when restricting ourself to (two copies of) the one-sided Witt algebra spanned by $\{ L_m, \bar L_m; m \leq 1 \}$ the situation is different. In that case there are only two possible embeddings with $n= \pm 1$ and by choosing $n=1$ the sum over $k$ in \eqref{comm} terminates after min$\{1-m, j\}$ terms instead of when $k=j$ which is not finite as we sum $j$ to infinity. As a consequence also the sum over $l$ terminates after $1-m$ terms and there appear only a finite number of generators and as we will see also only a finite number of terms.

In that case we proceed with the identity
\begin{align}\label{ide}
 \sum_{s_1 = k-1}^{j-1} ... \sum_{s_k = 0}^{s_{k-1}-1}= \frac{j!}{(j-k)! k!}  = {j \choose k} .
\end{align}
It is straightforward to see that it holds for $k=1$. Now assume that it holds for $k=k', 1 \leq k' < j$ and it follows that the lhs for $k=k'+1$ is
\begin{align}
\sum_{s'_1= k'}^{j-1} \sum_{s'_2 = k'-1}^{s'_1-1} ... \sum_{s'_{k'} = 0}^{s'_{k'-1} -1} = \sum_{s'_1= k'}^{j-1} {s'_1 \choose k'} = {j \choose k'+1} ,
\end{align}
where in the last step the hockey stick identity was used. Thus we proved \eqref{ide} by induction.
Therefore
\begin{align}
[\sigma_1, L_m]  =&  - \sum_{j = 0}^{\infty} \frac{(- \tilde a)^j}{j} \big[\left(L_n- \bar L_n \right)^{j}, L_m \big] \nonumber \\
=& L_m - \sum_{j = 1}^{\infty} \sum_{k=1}^{\text{min} \{j, 1-m\}}   \frac{(- \tilde a)^j}{j} {j \choose k} L_{m+k} (L_1 - \bar L_1)^{j-k} \prod_{p=0}^{k-1} (1-(m+p)) \label{sigcom}
\end{align}
and comparing with 
\begin{align}
\frac{d^k \sigma_1}{d L_1^k} = -\sum_{j = k}^{\infty} \frac{(- \tilde a)^j}{j} \frac{j!}{(j-k)!} (L_1 - \bar L_1)^{j-k}
\end{align}
one finds that the summands in \eqref{sigcom} and
\begin{align}\label{sigcom2}
L_m + \sum_{k=1}^{1-m} L_{m+k} \frac{d^k \sigma_1}{d L_1^k} \frac{1}{k!} \prod_{p=0}^{k-1} (1-(m+p))
\end{align}
are identical. Splitting the sums in \eqref{sigcom} and \eqref{sigcom2} according to
\begin{align*}
\left(\sum_{j=1}^{1-m} \sum_{k=1}^j + \sum_{j=2-m}^{\infty}\sum_{k=1}^{1-m} \right) ... \\
\left( \sum_{k=1}^{1-m}\sum_{j=k+1}^{1-m} + \sum_{k=1}^{1-m}\sum_{j=2-m}^{\infty} \right) ...
\end{align*}
and using
\begin{align}
\sum_{j =1}^{1-m} \sum_{k = 1}^j ... = \sum_{k=1}^{1-m} \sum_{j=k}^{1-m} ...
\end{align}
it follows that \eqref{comm} is indeed given by \eqref{sigcom2} if the one-sided Witt algebra with $n=1$ is considered.
Plugging the result into \eqref{delgen} yields
\begin{align}
\Delta_{\mathcal{F}_{J, 1}}(L_m) =& L_m \otimes \Pi_{+}^{m} + \sum_{l=1}^{1-m} \frac{1}{l!}\bigg(\sum_{k_1 = 0}^{1-m} \frac{1}{k!} \prod_{p_1=0}^{k_1-1} (1-(m+p_1)) \nonumber \\
&\times  \bigg( \sum_{k_2 = 0}^{1-(m+k_1)} \frac{1}{k_2!} \prod_{p_2=0}^{k_2-1} (1-(m+k_1 +p_2)) \nonumber \\
& \times \bigg(... \bigg( \sum_{k_l =0}^{1-(m+ k_1 + ... k_{l-1})} \frac{1}{k_l !} \prod_{p_l=0}^{k_l-1} (1-(m+k_1 +... + k_{l-1} +p_l)) A\otimes B \bigg)... \bigg), \label{delres}
\end{align}
where
\begin{align}
A \otimes B =& \left(\frac{L_0 + \bar L_0}{n} \right)^l \otimes L_{m+k_1 + ...+k_l} \frac{d^{k_1} \sigma_1}{d L_1^{k_1}}...  \frac{d^{k_l}\sigma_1}{d L_1^{k_l}} \\
 \frac{d^{k}\sigma_1}{d L_1^{k}} =& - (- \tilde a)^k \Pi_+^{-k} k!
\end{align}
and all sums are finite for $m \leq 1$. Similarly for $\bar L_m$ one finds
\begin{align}
\Delta_{\mathcal{F}_{J, 1}}(\bar L_m) =& \bar L_m \otimes \Pi_{+}^{m} + \sum_{l=1}^{1-m} \frac{1}{l!}\bigg(\sum_{k_1 = 0}^{1-m} \frac{1}{k!} \prod_{p_1=0}^{k_1-1} (1-(m+p_1)) \nonumber \\
&\times  \bigg( \sum_{k_2 = 0}^{1-(m+k_1)} \frac{1}{k_2!} \prod_{p_2=0}^{k_2-1} (1-(m+k_1 +p_2)) \nonumber \\
& \times \bigg(... \bigg( \sum_{k_l =0}^{1-(m+ k_1 + ... k_{l-1})} \frac{1}{k_l !} \prod_{p_l=0}^{k_l-1} (1-(m+k_1 +... + k_{l-1} +p_l)) \bar A\otimes \bar B \bigg)... \bigg),
\end{align}
where
\begin{align}
\bar A \otimes \bar B = \left(\frac{L_0 + \bar L_0}{n} \right)^l \otimes \bar L_{m+k_1 + ...+k_l} \frac{d^{k_1} \sigma_1}{d \bar L_1^{k_1}}...  \frac{d^{k_l}\sigma_1}{d \bar L_1^{k_l}}.
\end{align}

Then the antipodes follow from \eqref{antip} and read
\begin{align}
S_{\mathcal{F}_{\text{LC}}} (L_m) =& -  \sum_{l=1}^{1-m} \frac{1}{l!}\bigg(\sum_{k_1 = 0}^{1-m} \frac{1}{k!} \prod_{p_1=0}^{k_1-1} (1-(m+p_1)) \nonumber \\
&\times  \bigg( \sum_{k_2 = 0}^{1-(m+k_1)} \frac{1}{k_2!} \prod_{p_2=0}^{k_2-1} (1-(m+k_1 +p_2)) \nonumber \\
& \times \bigg(... \bigg( \sum_{k_l =0}^{1-(m+ k_1 + ... k_{l-1})} \frac{1}{k_l !} \prod_{p_l=0}^{k_l-1} (1-(m+k_1 +... + k_{l-1} +p_l)) S(A)\otimes  B \bigg)... \bigg) \Pi_+^{-m},
\end{align}
and $S(A)$ can be inferred from \eqref{SL0}
\begin{align}
S(A) = S ((L_0+\bar L_0)^l) = (-(L_0  + \bar L_0) \Pi_+)^l.
\end{align}

\subsection{Contraction Limit and Uniqueness of Deformations}

So far we obtained general Hopf algebras on the symmetry algebras for asymptotically AdS spacetimes which algebraically also carry over to the dS case easily. There are several reasons why the asymptotically flat case is of special interest though. One motivation is the possibility of deformed dispersion relations that is associated with non-trivial Hopf algebra structures but in (A)dS there are no true momenta. However, by performing the contraction limit $\Lambda \rightarrow 0$, one can obtain information about quantum groups in the three dimensional BMS from the $\Lambda-$BMS. 
For all r-matrices of the $\mathfrak{o}(4)$ the contraction limit was obtained in \cite{Kowalski-Glikman:2019ttm}. The resulting r-matrices were compared to the full classification of r-matrices of the Poincar\'e algebra in three dimensions from \cite{Stachura_1998} and it was claimed that all of them could be derived by an appropriate contraction limit. It should be noted that the contraction is ambiguous, i.e. the contraction of a class of r-matrices can be performed in different non-equivalent ways (cf. below) and not injective, i.e. there are r-matrices in the 3D Poincar\'e that can be obtained as a contraction from distinct $\mathfrak{o}(4)$ r-matrices. A general scheme describing corresponding classes of r-matrices and their contractions is depicted in Figure 1.

% However, as stated above, mapping to the set of all 3D Poincar\'e r-matrices by contraction is surjective. This also holds true when considering the subset of triangular r-matrices on $\mathfrak{o}(4)$ and the Poincar\'e algebra even though there is an example of a quasitriangular r-matrix that is contracted to a triangular one. 

Let us first explicitly perform the contraction limit of $r_1'$ from \eqref{classa}. To this end it is expressed in terms of $l_m, T_m$ with the help of \eqref{cwiso} and subsequently expanded in powers of $1/\sqrt{-\Lambda}$
\begin{align}
r_{1'} =& \frac{\alpha + \bar \alpha}{-\Lambda} (T_1 \wedge T_0 - T_{1}\wedge T_{-1} - T_{-1} \wedge T_0) + \nonumber \\
&\frac{\beta}{\sqrt{-\Lambda}}  (l_1 +l_{-1} + 2l_0)\wedge (T_1 +T_{-1} + 2T_0) + \frac{\alpha}{\sqrt{-\Lambda}} (...) + \beta(...) .
\end{align}
In order to obtain a finite result we have to rescale $(\alpha + \bar \alpha) \rightarrow (\hat \alpha + \hat {\bar \alpha})(-\Lambda)$ and $\beta \rightarrow \hat \beta \sqrt{-\Lambda}$. Then taking the limit $\Lambda \rightarrow 0$ yields
\begin{align}
\hat r_{1',a} = (\hat \alpha + \hat{\bar \alpha}) (T_1 \wedge T_0 + T_{1} \wedge T_{-1}  - T_{-1} \wedge T_0) + \hat \beta (l_1 +l_{-1} + 2l_0)\wedge (T_1 +T_{-1} + 2T_0).
\end{align}
This is not the only possibility to abtain a finite limit, in the case $\alpha =-\bar \alpha$ one can also rescale $\alpha \rightarrow \hat \alpha \sqrt{-\Lambda}$ to get
\begin{align}
\hat r_{1',b} =& \beta (l_1 \wedge T_1 + l_{-1} \wedge T_{-1} + 4 l_0 \wedge T_0) + (2 \beta + \alpha)(l_1 \wedge T_0 + l_0 \wedge T_{-1}) \nonumber \\
&+ (2 \beta  - \alpha) (l_0 \wedge T_1 + l_{-1} \wedge T_0) + (\beta + \alpha) l_1 \wedge T_{-1} + (\beta - \alpha) l_{-1} \wedge T_1.
\end{align}
Similarly the contraction limit can be performed for all r-matrices in \eqref{classa}-\eqref{classb}.
Comparing to (the triangular part of) the classification of r-matrices on $\mathfrak{P}_3$ by \cite{Stachura_1998} shows that the contractions of $r_{1'}$ are in general not automorphic to it via a $\mathfrak{B}_3$ automorphism. Instead the set of r-matrices of the $\mathfrak{P}_3$ up to $\text{Aut}(\mathfrak{B}_3)$ is strictly larger than up to $\text{Aut}(\mathfrak{P}_3)$ similar to the case of non-vanishing cosmological constant.

It is also not clear a priori if the contraction from \eqref{classa}-\eqref{classb} to this set is surjective and one has to take into account that the contraction limit can be performed along different axis as explained in the following. The (anti) de Sitter algebra \eqref{ads3}-\eqref{ads3l} is isomorphic to 
\begin{align}
[M_{AB}, M_{CD}] = \delta_{AC} M_{BD} - \delta_{BC} M_{AD} + \delta_{BD} M_{AC} - \delta_{AD} M_{BC},
\end{align}
where the indices range from $1$ to $4$ and 
\begin{align}
M_{+-} & = M_{13}, \quad M_{+2} = M_{12} + M_{32}, \quad M_{-2} = M_{12}- M_{32}, \\
K_{\pm} & = M_{14} \pm M_{34}, \quad K_2 = M_{24},
\end{align}
i.e. the fourth axis is chosen for contraction. With the isomorphism 
\begin{align}
H = & -\frac{i}{2} (M_{12} + M_{34}), \quad \bar H = \frac{i}{2} (M_{12} - M_{34}), \\
E_{\pm} = & -\frac{i}{2} (M_{23} + M_{14}) \mp \frac{1}{2}(M_{24} - M_{13}),\\
\bar E_{\pm} = & \frac{i}{2} (M_{23} - M_{14}) \mp \frac{1}{2}(-M_{24} - M_{13})
\end{align}
one can express the r-matrices in terms of $M_{AB}$. Depending on which axis is chosen the result of the contraction differs, e.g. when choosing the second instead of the fourth axis
\begin{align}
K_{i} = M_{i2}, \quad J_{i} = \epsilon_{ijk} M_{jk}
\end{align}
one finds for 
\begin{align}
r = (E_+ + \bar E_+) \wedge (H + \bar H),
\end{align}
which is automorphic to $r_I$, 
\begin{align}
r = \frac{i}{2} (i J_3 + J_4) \wedge J_1
\end{align}
where the contraction limit can be taken without rescaling. The $J_i$ satisfy 
\begin{align}
[J_1, J_3] = J_4, \quad [J_1, J_4] = J_3, \quad [J_3, J_4] = J_1
\end{align}
and can thus be related to the three dimensional Lorentz sector via
\begin{align}
J_3 = -l_0, \quad J_1 = \frac{l_1+l_{-1}}{2},\quad J_4 = \frac{l_1-l_{-1}}{2}
\end{align}
leading to 
\begin{align}
\hat r = l_1 \wedge l_{-1} + l_0 \wedge(l_1 + l_{-1}) .
\end{align}
Note that this r-matrix of the $\mathfrak{P}_3$ can not be obtained from a contraction of \eqref{classa}-\eqref{classb} which are associated with the fourth axis.

But even after taking into account the possibility to contract along different axis it turns out that there are triangular r-matrices in $\mathfrak{P}_3$, e.g.
\begin{align}
r = l_0 \wedge T_1 + \Theta_1 T_1 \wedge T_0 + \Theta_2 T_1 \wedge T_{-1}
\end{align}
that can not be obtained in its general from a contraction of a triangular r-matrix. This is important insofar it would enable a constructive method to obtain a twist for all deformations as we will see now.

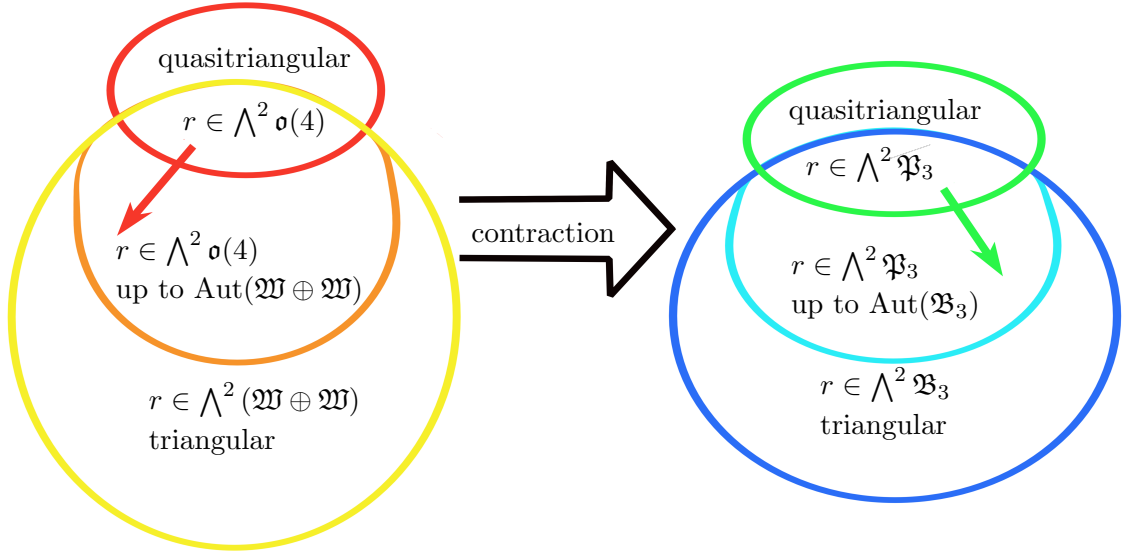
\begin{figure}
\centering
\begin{tikzpicture}
    \draw (0, 0) node[inner sep=0] {\includegraphics[width=16cm]{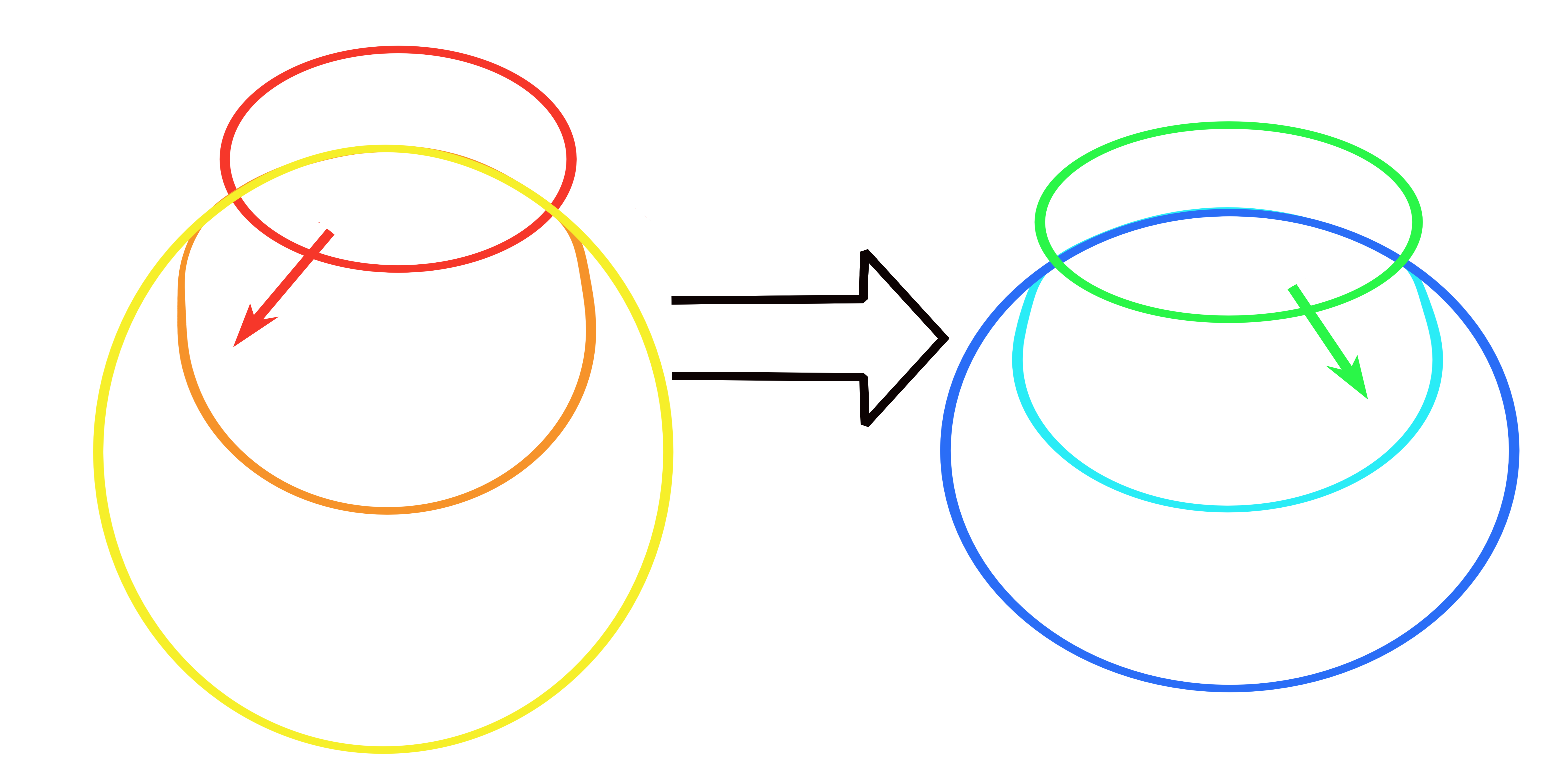}};
    \draw (-3.8, 2.0) node {$r \in \bigwedge^2 \mathfrak{o}(4)$};
	\draw (-3.8, 2.8) node {quasitriangular};
	\draw (4.5, 2.1) node {quasitriangular};
	\draw (0, 0.5) node {contraction};
	\draw (4.3, 1.4) node {$r \in \bigwedge^2 \mathfrak{P}_3$};
    \draw (-4.0, 0.0) node {\begin{tabular}{l}
    $r \in \bigwedge^2 \mathfrak{o}(4)$  \\
	up to $\text{Aut}(\mathfrak{W}\oplus \mathfrak{W})$
    \end{tabular}};
    \draw (-3.8, -2.0) node {\begin{tabular}{l}
     $r \in \bigwedge^2 \left(\mathfrak{W}\oplus \mathfrak{W}\right)$ \\
     triangular 
     \end{tabular}};
	\draw (4.5, -1.8) node {\begin{tabular}{l}
	$r \in \bigwedge^2 \mathfrak{B}_3$ \\
    triangular 
     \end{tabular}};	
	\draw (4.5, -0.2) node {\begin{tabular}{l}
    $r \in \bigwedge^2 \mathfrak{P}_3$  \\
	up to $\text{Aut}(\mathfrak{B}_3)$
    \end{tabular}};
\end{tikzpicture}
\caption{Schematic depiction of r-matrices in three dimensional asymptotic symmetry algebras. The red arrow represents the quotient $\text{Aut}(\mathfrak{o}(4))/\text{Aut}(\mathfrak{W} \oplus \mathfrak{W})$ and the green arrow $\text{Aut}(\mathfrak{P}_3)/\text{Aut}(\mathfrak{B}_3)$. All the r-matrices inside the yellow and blue circle are triangular.}
\end{figure}

Namely there is the possibility in performing the contraction limit on the level of the full twist. As an example let us consider the light-cone twist 
\begin{align}\label{lctw}
\mathcal{F}_{\text{LC}} = e^{ L_0 \otimes \log \left( 1+ a L_n \right)} e^{ \bar L_0 \otimes \log \left( 1 - a \bar L_n \right)} 
\end{align}
corresponding also to \eqref{lcr2}. We express the $L_m, \bar{L}_m$ in terms of $l_m, T_m$
 \begin{align}
 \mathcal{F}_{\text{LC}} = & \exp \left( \frac{1}{2} \left(l_0 + \frac{T_0}{\sqrt{-\Lambda}}\right) \otimes \log  \left(1 + \frac{a}{2}\left(l_n + \frac{T_n}{\sqrt{-\Lambda}} \right) \right) \right)  \nonumber \\
 & \times \exp \left( \frac{1}{2} \left(l_0 - \frac{T_0}{\sqrt{-\Lambda}}\right) \otimes \log  \left(1 - \frac{a}{2}\left(l_n - \frac{T_n}{\sqrt{-\Lambda}} \right) \right) \right) . \label{lctwist}
 \end{align}

In order to obtain a finite contraction limit we have to rescale $a \rightarrow a' = a/\sqrt{-\Lambda}$. Expanding the first exponent of the twist in powers of $\sqrt{-\Lambda}$ and taking the limit then results in
\begin{align}
\frac{1}{2} \left( l_0 + \frac{T_0}{\sqrt{\Lambda}} \right) & \otimes \sum_{j=1} \frac{- (-a')^j}{j} \left( \sqrt{-\Lambda}l_n + T_n\right) \nonumber \\
= & \frac{1}{2} l_0 \otimes \log ( 1 + a' T_n) + \frac{1}{2} T_0 \otimes \sum_{j=1} \frac{-(-a')^j}{j} j l_n T_n^{j-1} + \mathcal{O}( \sqrt{-\Lambda}) \nonumber \\
= & \frac{1}{2} l_0 \otimes \log ( 1 + a' T_n) + \frac{1}{2} T_0 \otimes l_n a'\left( 1 + a' T_n\right)^{-1}.
\end{align}
After repeating this procedure for the second exponent the final twist is given by
\begin{align}\label{lc2}
\mathcal{F}_{\text{LC}} = \exp \left( l_0 \otimes \log (1 + a' T_n) + a' T_0 \otimes  l_n \left( 1+ a' T_n \right) \right).
\end{align}
It automotically satisfies the 2-cocycle condition in $\mathfrak{B}_3$ since the twist \eqref{lctwist} satisfies it in $\mathfrak{W}\oplus \mathfrak{W}$.

Similarly from the contraction of the abelian twist discussed in section \ref{sec411} we obtain the abelian twist discussed in \cite{Borowiec:2020ddg}, where its physical interpretation is discussed and compared to the Jordanian twist deformation. 

%  Therefore, to obtain a finite contraction limit one can rescale $T_m \rightarrow  \sqrt{-\Lambda}T_m$ (which is an automorphism of the BMS) and $a \rightarrow a' = \frac{a}{\Lambda}$. We already know that $\mathcal{F}_{\text{LC}}$ satisfies the 2-cocycle condition in $\mathfrak{W} \oplus \mathfrak{W}$ and it follows that the contracted twist fulfills the condition in $\mathfrak{B}_3$. This can be seen because one can introduce e.g. a normal ordering of the supertranslations in \eqref{coc-cond} and write it as a power series in $a$ where each coefficient has to satisfy the relation separately. Then, after rescaling and taking the limit $\Lambda \rightarrow 0$ (implying that the commutator of the $T_m$ vanishes) \eqref{coc-cond} still holds and we obtained a valid twist for$\mathfrak{B}_3$
% \begin{align}\label{lc2}
%  \mathcal{F}_{\text{LC}} =& \exp \left( \frac{1}{2}  (l_0 + T_0) \otimes \log \left( 1 + \frac{a'}{2} (l_n + T_n) \right) \right) \nonumber \\
%  & \times  \exp \left( \frac{1}{2}  (l_0 -T_0) \otimes \log \left( 1 - \frac{a'}{2} (l_n - T_n) \right) \right).
%  \end{align}
 
\subsubsection*{(Non-)uniqueness of the Twist} 
From \cite{Borowiec:2020ddg} we also know that the extended Jordanian twist of the form 
\begin{align}
 \mathcal{F}_{\text{eJ}} = \exp \left( \frac{a''}{2} l_n \otimes T_0 \right) \exp \left( - \frac{l_0}{n} \otimes \log \left(1 + a'' n T_n \right) \right)
\end{align}
exists for the $\mathfrak{B}_3$ r-matrix $r=l_0 \wedge T_n + l_n \wedge T_0$ which is the contraction limit of \eqref{lcr}. Comparing the two twists reveals that they are related by a flip in first order and differ in higher orders. However, the inequivalence is only superficial as we have to take into account automorphisms on the universal envelope. We find there exist invertible elements $\omega \in U \mathfrak{B}_3 [[1/\kappa]]$ inducing the automorphisms  
\begin{align}
f(l_m) = \omega^{-1} l_m \omega, \quad f (T_m) = \omega^{-1} T_m \omega
 \end{align}
 by a similarity transformation. In general for every twist deformed Hopf algebra with $\mathcal{F}$ one can obtain a gauge equivalent twist via $\mathcal{F}_{\omega} = \omega^{-1} \otimes \omega^{-1} \mathcal{F} \Delta (\omega)$. The new twist then satisifies the 2-cocycle condition because
 \begin{align}
 {\mathcal{F}_{\omega}}_{12} ( \Delta \otimes 1) ( \mathcal{F}_{\omega}) &= (\omega^{-1} \otimes \omega^{-1}) F_{12} (\Delta(\omega) \otimes \omega) \Delta( \omega^{-1} ) \otimes \omega^{-1}) (\Delta \otimes 1) \mathcal{F} (\Delta(\omega) \otimes \omega) \nonumber \\
 &=  (\omega^{-1} \otimes \omega^{-1}) \mathcal{F} _{23} ( 1 \otimes  \Delta) \mathcal{F}  (\Delta (\omega) \otimes \omega) \nonumber \\
 &= {\mathcal{F}_{\omega}}_{23} (1 \otimes \Delta) \mathcal{F}_{\omega} 
 \end{align}
 and $f(x) = \omega x \omega^{-1}$ establishes the isomorphism between the twisted Hopf algebras
 \begin{align}
 \Delta_{\mathcal{F} } \circ f = (f \otimes f) \circ \Delta_{\mathcal{F}_{\omega}}.
 \end{align}
 If the untwisted Hopf algbebra admits a $*$-structure the twist has to satisfy
 \begin{align}
 \mathcal{F}^{* \otimes *} = \mathcal{F}^{-1},
 \end{align}
 i.e. be {unitary} in order to preserve the $*$-structure. On the invertible element $\omega$ this enforces the unitarity condition as well
 \begin{align}
 \mathcal{F}_{\omega}^{* \otimes *}  &= \Delta (\omega^*) \mathcal{F}^{-1} ( {\omega^*}^{-1} \otimes  {\omega^*}^{-1}) \overset{!}{=}  \mathcal{F}_{\omega}^{-1} = \Delta (\omega^{-1}) \mathcal{F}^{-1} (\omega \otimes \omega) \\
 \Leftrightarrow \omega^* &= \omega^{-1}.
 \end{align}
 
 In our particular example we find that in first order of $1/\kappa$ the isomorphism induced from the element 
 \begin{align}
 \omega = e^{- \frac{a'}{4} (l_n  T_0 + T_0 l_n)}
 \end{align}
 relates the extended Jordanian and the contraction limit of the light-cone twist for $a'' = - a'$. It is easy to see that it is hermitian with respect to the reality condition $l_m^* = -l_m, T_m^* = -T_m$ and $a' \in i \mathbb{R}$.

\subsection{Deformations of the Surface Charge Algebra}

As we noted in section \ref{secsc} the algebra of the surface charges in an asymptotically AdS spacetime differs from the previously examined $\mathfrak{W} \oplus \mathfrak{W}$ by a central extension. Therefore we want to investigate whether this has any impact on the possible deformations. 

First let us note without proof that the first cohomology group $H^1(\mathfrak{Vir} \oplus \mathfrak{Vir}, \bigwedge \mathfrak{Vir} \oplus \mathfrak{W})$ vanishes and there are again no ad-invariant elements in the exterior product. As a consequence all LBA are coboundary and we will now investigate how the classification of the r-matrices is affected. Note that there is no automorphism on $\mathfrak{Vir} \oplus \mathfrak{Vir}$ that mixes elements of $\mathfrak{W}$ with central elements so it is enough to consider \eqref{classa}-\eqref{classb}. {More exactly, the formulas} \eqref{au1} are still valid if we assume $\phi(c_L)=c_L$.
By a straightforward computation one finds that all of these r-matrices are still triangular except for those containing $a_1$ if we choose a different embedding. In particular for $a_1$ with $L_{\pm 1} \rightarrow L_{\pm n}$ we have
\begin{align}
    [[a_1, a_1]] = \frac{(n^3-n)}{6} (L_n \wedge c_L \wedge L_{-n} + \wedge c_L \wedge L_0 + L_{-n} \wedge c_L \wedge L_0),
\end{align}
so it no longer defines a LBA.

Furthermore it is easy to see that all $r \in c_L \bigwedge \left( \mathfrak{W} \oplus \mathfrak{W}\right)$ are valid r-matrices. By calculating the Schouten brackets of combinations with r-matrices from \eqref{classa}-\eqref{classb} only r-matrices of the form
\begin{align}\label{crm1}
    r = c_L \wedge L_p + L_p \wedge \left(\sum_q \bar L_q \right), \\
    r = c_L \wedge L_p + \bar r, \\
    r = c_L \wedge L_p + \alpha L_p \wedge L_0\label{crm2}
\end{align}
are possible, where $\bar r$ is a triangular r-matrix generated by $\bar L_q$.

The central extension thus can impact the LBAs. For example with $r = c_L \wedge L_p$ we obtain
\begin{align}
    \delta_r (L_m) = c_L \wedge (m-p) L_{m+p}, \quad \delta_r(\bar L_m) = 0.
\end{align}
But also Lie bialgebras from r-matrices that contain no central element can contribute extra terms, e.g. $r = \chi L_0 \wedge L_n$ yields
\begin{align}
    \delta_r(L_{-n}) = -n \chi L_{-n} \wedge L_n - \chi \frac{(n^3-n)}{12} L_0 \wedge c_L.
\end{align}
For all r-matrices from \eqref{crm1}-\eqref{crm2} we can write down the twist. This is easy to see in the first two cases as everything is abelian. In the last case note that we can obtain the twist from the Jordanian twist by simply redefining $L_0 \rightarrow L_0 - \alpha c_L$ which leaves $[L_0, L_p]$ invariant.

From the twists we can directly compute the coalgebra structures. In general these will contain infinite expressions which, as we will explore in the next section, could be remedied by considering one-sided algebras. Even though this works in the case of the Witt algebra, with a central extension the one-sided algebra is pointless to consider as the central elements do not appear in the algebra sector. Thus out of \eqref{crm1}-\eqref{crm2} only $r =  c_L \wedge (\chi L_0 + \bar \chi \bar L_0)$ leads to a finite coalgebra sector. In particular,
\begin{align}
    \Delta (L_m) = L_m \otimes 1 + \exp ( -m \chi c_L) \otimes L_m
\end{align}

\section{One-sided Witt Algebra and Specialization}

So far the Hopf algebras we considered were defined with the $h$-adic topology and thus allowed for infinite power series in the formal parameter $1/\kappa$. While this is mathematically consistent it is ultimately problematic when interpreting the formalism in a physical context where $1/\kappa$ is to be identified with an energy scale of the order of the Planck mass.  The problem of finding a Hopf algebra (the so-called q-analog) with the same (co)algebra structure where the formal parameter can be specialized to a complex (or real) parameter is known as specialization \cite{klimyk2011quantum, Borowiec:2014aqa}. Most importantly, all the structures in the q-analog need to be finite power series in the generators. 

Let us study the specialization on the examples of the abelian and the Jordanian twist respectively. For the abelian twist the formulas \eqref{copab1}-\eqref{copab2} show that only a finite number of generators appear in the coproducts but there are infinite power series in $a \equiv \frac{im}{\kappa n^2}$. Thus the full $\mathfrak{W} \oplus \mathfrak{W}$ can be turned into the q-analog by adding the elements 
\begin{align}
e^{a L_0} \equiv K, \quad e^{-aL_0}  \equiv K^{-1}, \quad e^{a \bar L_0} \equiv \bar K, \quad e^{-a \bar L_0}  \equiv \bar K^{-1},
\end{align}
to the algebra. Furthermore, define $q = e^{a}$ and the extra commutation relations become
\begin{align}
e^{a L_0 } L_m e^{-a L_0 } =& \sum_{j=0}^{\infty} \frac{a^j}{j!} [L_0 , [...,[ L_0 , L_m]...] = e^{-am} L_m = q^{-m} L_m \\
\Rightarrow [K, L_m] = & q^{-m} L_m e^{aL_0 } - L_m e^{a L_0 } = (q^{-m} -1) L_m K
\end{align}
and similarly
\begin{align}
[K^{-1}, L_m] =& (q^m -1)L_m K^{-1}, \quad [K, \bar L_m] = 0, \\
[K^{-1}, \bar L_m] = & 0, \quad [\bar K, \bar L_m] = (q^{-m} -1) \bar L_m \bar K\\
[\bar K^{-1}, L_m] = & 0, \quad [\bar K^{-1}, \bar L_m] = (q^m-1) \bar L_m \bar K^{-1}, \\
[K, K^{-1}] = & [K, \bar K] = [K, \bar K^{-1}] =0.  
\end{align}
It is also easy to compute
\begin{align}
\Delta_{\mathcal{F}_A}(K) = & K \otimes K, \quad \Delta_{\mathcal{F}_A}(K^{-1}) = K^{-1} \otimes K^{-1}, \\
S_{\mathcal{F}_A}(K) = & -K , \quad S_{\mathcal{F}_A}(K^{-1}) = - K^{-1}
\end{align}
and reexpressing \eqref{copab1}-\eqref{copab2} gives
\begin{align}
\Delta_{\mathcal{F}_A}(L_m) = & K^m \bar K^m \otimes L_m + L_m \otimes K^{-m} \bar K^m  \\
\Delta_{\mathcal{F}_A}(\bar L_m) = & K^{-m} \bar K^{-m} \otimes \bar L_m + \bar L_m \otimes K^{m} \bar K^{-m} .
\end{align}
Endowed with this algebra and coalgebra structures the set of polynomials in the generators $\{ L_m, \bar L_m, K, K^{-1}, \bar K, \bar K^{-1}\}$ does indeed form a q-analog of the twisted Hopf algebra and it can be defined for any $q \in \mathbb{C}$. In particular the classical limit $\kappa \rightarrow \infty \leftrightarrow q \rightarrow 1$ gives simply the Lie algebra $\mathfrak{W} \oplus \mathfrak{W}$ but extended by the central elements $K, \bar K$.

In the case of the Jordanian twist the situation is different. We discovered in \eqref{delgen} that for $L_m, m \in \mathbb{Z}$ the coproduct contains infinitely many different generators and thus it would be impossible to define a q-analog. However, by restricting to two copies of the one-sided Witt algebra $\mathfrak{W}_-$ containing $L_m, m \leq 1$ it was shown that all coproducts contain finitely many terms. Similarly one could use the embedding corresponding to $n=-1$ and restrict to $\mathfrak{W}_+$ containing $L_m, m \geq -1$. In order to express all algebra and coalgebra relations involving only finite powers of $1/\kappa$ the elements $\Pi_+$, defined in \eqref{defpi} and its inverse $\Pi_+^{-1}$ are used. The additional commutation relations then read
\begin{align}
[\Pi_+, L_m] =& \tilde a (1-m) L_{m+1}, \\
[\Pi_+^{-1}, L_m] =& \sum_{j = 0}^{\infty} \frac{(-1)^{\underline{j}}}{j!}  \tilde a^j [(L_1 - \bar L_1)^j, L_m] \nonumber \\
=& \sum_{j=0}^{\infty} \sum_{k=1}^{\text{min}\{1-m, j\}} \frac{(-1)^{\underline{j}}}{j!} {j \choose k} L_{m+k} \tilde a^j (L_1 - \bar L_1)^{j-k} \left( \prod_{r = 0}^{k-1} (1-m-r) \right) \nonumber \\
=& \sum_{k=1}^{1-m} L_{m+k} \frac{d^k e^{-\sigma_1}}{d L_1^k} \left( \prod_{r = 0}^{k-1} (1-m-r) \right) \nonumber \\
= & \sum_{k=1}^{1-m} \frac{(-1)^{\underline{k}}}{k!} \tilde a^k L_{m+k} \Pi_+^{k-1} \left( \prod_{r = 0}^{k-1} (1-m-r) \right) 
\end{align}
and similarly
\begin{align}
[\Pi_+, \bar L_m] =& -\tilde a (1-m) \bar L_{m+1}, \\
[\Pi_+^{-1}, \bar L_m] = &  \sum_{k=1}^{1-m} \frac{(-1)^{\underline{k}}}{k!} (-\tilde a)^k \bar L_{m+k} \Pi_+^{k-1} \left( \prod_{r = 0}^{k-1} (1-m-r) \right) .
\end{align}
From \eqref{delres} one has in particular
\begin{align}
\Delta_{\mathcal{F}_J} (L_1) = L_1 \otimes \Pi_+ + \otimes L_1, \quad \Delta_{\mathcal{F}_J} (L_1) = \bar L_1 \otimes \Pi_+ + \otimes \bar L_1,
\end{align}
leading to
\begin{align}
\Delta_{\mathcal{F}_J} (\Pi_+) & = \Pi_+ \otimes \Pi_+, \quad \Delta_{\mathcal{F}_J} (\Pi_+^{-1})  = \Pi_+^{-1} \otimes \Pi_+^{-1} , \\
S_{\mathcal{F}_J} (\Pi_+) & = -\Pi_+, \quad S_{\mathcal{F}_J} (\Pi_+^{-1}) = - \Pi_+^{-1}. 
\end{align}
All these formulas are well defined for $\tilde a \in \mathbb{C}$ and for $\kappa \rightarrow \infty$ the elements $\Pi_+, \Pi_+^{-1}$ become central. Thus, similar to the abelian twist, the classical limit is the centrally extended Lie algebra $\mathfrak{W}_+ \oplus \mathfrak{W}_+$.

Additionally, if we consider the centrally extended algebra of surface charges, the r-matrix $r =  c_L \wedge (\chi L_0 + \bar \chi \bar L_0)$ with the twist $\mathfrak{F} = \exp(  c_L \otimes (\chi L_0 + \bar \chi \bar L_0))$ induces a Hopf algebra which admits a specialization by simply adding the elements $\Pi = \exp (\chi c_L), \bar \Pi = \exp (\bar \chi c_L)$. These elements are thus just redefinitions of the Brown-Henneaux central charge.

It turns out that all twist deformations except for the abelian twist do not have a q-analog on the full Witt algebras. But those (and only those) which do not contain both $L_1$ and $L_{-1}$ or $\bar L_{1}$ and $\bar L_{-1}$ simultaneously can be shown to permit specialization on the one-sided Witt algebras in a similar way as for the Jordanian twist. Therefore we will investigate what physical implications the restriction of the generators has. 

To this end we consider an asymptotically flat spacetime in four dimensions, described by $\mathfrak{B}_4$ (see also \cite{Borowiec:2020ddg}).
Recall that the superrotation Killing vectors are parametrized by functions $R^A$ on the sphere. For $m \geq -1$ these functions do not contain negative powers of $z, \bar z$ and are thus holomorphic on the whole sphere except for $z = \infty$. Note that only the ordinary rotations with $m=0,1,2$ are globally defined as the vectorfields $R^z_m \equiv z^m \partial_z, m<0$ and, after redefining $\omega = z^{-1}$,$R^z_m = \omega^{2-m} \partial_{\omega}, m>2$  have a singularity at the origin \cite{Compere:2018aar}. 

In contrast, consider the following construction due to Penrose where Minkowski space is cut along the light-cone $u=0$ \cite{Penrose:1976},\cite{Strominger:2016wns}. Then, after performing a diffeomeorphism on the $u>0$ patch, it is glued together such that the metric is continuous at $u=0$. That procedure introduces singularities and was later linked to cosmic strings \cite{Gleiser:1989vt}. A cosmic string is a topological defect with dimension one and is conjectured to exist if in the early universe the topology was not simply connected. The geometry containing a cosmic string is not exactly asymptotically flat because of the singularities but it satisfies a weaker requirement and is said to be asymptotically locally flat. A snapping string with ends at $z=0, \infty$ that starts to snap from $u=0$ is indeed described by Penrose' construction and furthermore one can show that certain superrotations of flat space yield cosmic strings. In other words a superrotation that is only meromorphic, i.e. isolated singularities are allowed, maps a flat geometry to a flat geometry except at the singularities \cite{Strominger:2016wns}. 

If the results we obtained in three dimensions carry over qualitatively to the four dimensional case, i.e. that the consistent specialization of particular twist deformations requires the restriction to the one-sided $\mathfrak{B}_{4+}$, then we could conclude that the remaining superrotations do not allow for the formation or decay of cosmic strings. Thus phenomenological evidence for the existence of cosmic strings, e.g. from observing gravitational wave signatures of their decay, could be used to constrain theories of quantum groups and non-commutative geometry.

\section{Conclusion}

It was shown in this work that all Lie bialgebra structures on the symmetry algebra of asymptotically (A)dS spacetime in 3 dimensions are coboundary and triangular and can thus be quantized with the help of the Drinfeld twist technique. Physically viable r-matrices, that is those which are compatible with singling out an embedding representing a vacuum choice, are all classified. Also the triangularity condition constrains the possible Lie bialgebras and in particular some of the structures that are defined on the 3-dimensional Poincar\'e algebra related to $\kappa$-Poincar\'e quantum groups are eliminated due to this.
With the help of the quantization of the Lie bialgebra structures on (real forms of) $\mathfrak{o}(4, \mathbb{C})$ there is a constructive way to obtain the associated Hopf algebras in all orders of the deformation parameter also for the revised classes of r-matrices in the infinite-dimensional $\mathfrak{W} \oplus \mathfrak{W}$ algebra and its BMS$_3$ contraction  limits, cf. a schematic overview presented in Figure 1. When performing this twist procedure it becomes apparent that the specialization of the formal deformation parameter to real values can not be done for all Hopf algebras. Rather, in these cases, this is only possible when a subalgebra of the asymptotic symmetry algebra is considered. We propose that this would have testable consequences when transferred to a realistic setting, namely the existence of cosmic strings would be inconsistent with the quantum group deformations. Further phenomenological consequences were already studied for the flat case in \cite{Borowiec:2020ddg} and we make contact with this work by performing a contraction limit. 

There is a number of problems that we hope to be able to address in a future. First of all it would be of great interest to directly derive the deformation of 3 dimensional BMS algebra  by using the non-perturbative methods similar to those that made it possible to derive the deformation of quantum AdS algebra of charges in \cite{Cianfrani:2016ogm}. Given that by AdS$_3$/CFT$_2$ asymptotic symmetries of gravity with negative cosmological constant should correspond to the symmetries of the conformal field theory on the asymptotic boundary of spacetime it is natural to investigate CFTs with deformed conformal symmetries, to understand, among others, what would be the origin of the deformation, corresponding to the non-perturbative quantum gravity effect leading to the deformation of symmetries in the bulk.

In this paper we considered only deformations of the   $\Lambda$-BMS$_3$ algebra \eqref{3dbms}-\eqref{3dalg}. Recently,  various generalizations of this algebra were proposed, which result from boundary conditions different from the Brown-Henneaux ones \cite{Brown:1986nw} adopted here. In the paper \cite{Compere:2013bya} the authors consider  chiral boundary conditions and the resulting algebra of charges differs from  \eqref{3dbms}-\eqref{3dalg}. It would be of interest to look for possible deformations of this algebra, however since the translational sector of it differs from the one we consider, to do so one has to adopt a new class of twists. {Another generalization of the $\Lambda$-BMS$_3$ algebra considered here was proposed recently in} \cite{Fuentealba:2020zkf} {where the conformal extension of BMS algebra were considered. Since $\mathfrak{o}(3,1)$ is a subalgebra of the three dimensional conformal algebra $\mathfrak{o}(3,2)$, to deform the latter one can readily use the twists constructed here. In the case of the conformal BMS$_3$ there are also other twist deformations, which it would be of interest to investigate in some details
One should notice however that in contrast to $\mathfrak{o}(4,\mathbb{C})$ case not all quantum deformations of $\mathfrak{o}(5,\mathbb{C})$ algebra are yet known. Other proposals leading to different algebras that can be, in principle, deformed using the method presented in this paper have been reported in the papers} \cite{Afshar:2016wfy}, \cite{Grumiller:2019fmp}, \cite{Batlle:2020hia}, \cite{Adami:2020ugu}.

In this paper we considered only the 3-dimensional model of non-zero cosmological constant $\Lambda$-BMS algebra. The reason why we choose to investigate here the simpler 3-dimensional $\Lambda$-BMS algebra is that in 4 dimensions the non-vanishing cosmological constant extension of the BMS algebra has the structure of a Lie algebroid, with structure functions instead of structure constants \cite{Safari:2019zmc}, \cite{Compere:2019bua}, \cite{Compere:2020lrt}, \cite{Barnich17}. Despite some attempts to generalize the notions of quantum groups/Lie bialgebras to bialgebroids/Hopf algebroids \cite{Lu96}, \cite{Brzezinski:2002}, \cite{Pachol17} there is no established concept for deformations of Lie algebroids.

\section*{Acknowledgments}
This work is supported by funds provided by the Polish National Science Center (NCN),
project UMO-2017/27/B/ST2/01902 and for  JKG, and JU also by the project number
2019/33/B/ST2/00050.

\appendix
\section{Proof of the Cohomology Theorems}\label{appA}

\subsection{Proof of Theorem 1}

We start by noting that the 1-cocycles $\delta$ can be separated by their degree $d \in \mathbb{Z}$. This degree is derived from the grading of $\mathfrak{W}$, i.e.
\begin{align}
\delta (L_m) = L_i \wedge L_j
\end{align}
has degree $d = i+j-m$. The separation by degree follows from the fact that a cocycle which, applied to elements of $\mathfrak{W}$, results in terms with different degree can be split into cocycles of homogeneous degree which have to fulfill the cocycle condition \eqref{coco} independently.
Let us first consider cocycles of degree $d \neq 0$. We will show that all such cocycles $\delta$ are cohomolog to $0$, i.e. that $\delta'(L_m) = \delta(L_m) - (\partial_0 r)(L_m) = 0$ for all $m \in \mathbb{Z}$. 
Let $\delta$ be a cocycle of homogenous degree $d$ such that
\begin{align}
\delta (L_m) = \sum_{i_m, j_m \in I_m} \alpha^m_{i_m j_m} L_{i_m} \wedge L_{j_m}, 
\end{align}
where $\alpha^m_{i_m j_m} \in \mathbb{R}; i_m, j_m \in \mathbb{Z}$ and $I_m$ are finite subsets of $\mathbb{Z}$.
Choose a 0-cochain
\begin{align}
r = -\sum_{i_0, j_0 \in I_0} \frac{\alpha^0_{i_0 j_0}}{i_0 + j_0} L_{i_0} \wedge L_{j_0}.
\end{align}
 Then we have
\begin{align}
\delta'(L_0) &= \sum_{i_0, j_0 \in I_0} \alpha^0_{i_0, j_0} L_{i_0} \wedge L_{j_0} - [L_0 \otimes 1 + 1 \otimes L_0, r] \nonumber \\
&= \sum_{i_0, j_0 \in I_0} \alpha^0_{i_0, j_0} L_{i_0} \wedge L_{j_0} - \sum_{i_0, j_0 \in I_0} \frac{\alpha^0_{i_0 j_0}}{i_0 + j_0} (i_0 + j_0) L_{i_0} \wedge L_{j_0} = 0.
\end{align}
From the cocycle condition
\begin{align}
\delta'([L_0, L_m]) = [L_0 \otimes 1 + 1 \otimes L_0, \delta'(L_m)] - [L_m \otimes 1 + 1 \otimes L_m, \delta'(L_0)],
\end{align}
for $m \neq 0$, we infer
\begin{align}
m \delta(L_m) &= \sum_{i_m, j_m \in I_m} (i_m + j_m) \alpha^m_{i_m j_m} L_{i_m} \wedge L_{j_m} \nonumber \\
&= (d+m) \delta'(L_m) \\
\Rightarrow  \delta'(L_m) &= 0,
\end{align}
which concludes the proof for cocycles of degree $d \neq 0$.

Next, let us consider cocycles of degree $d=0$ which can be written in the form
\begin{align}
\delta(L_m) = \sum_{i_m \in I} \gamma^m_{i_m} L_{m - i_m}\wedge L_{i_m}.
\end{align}
Note that without loss of generality we can restrict the indices $i_m$ to be smaller than $m/2$ since otherwise, i.e. if there is an index $i_m > m/2$, we simply substitute $i'_m = m-i_m$ and ${\gamma'}^m_{i'_m} = \gamma^m_{i_m}-\gamma^m_{im-m}$ to describe the same cocycle. We will make repeated use of this in the rest of the proof. 

The conditions
\begin{align}
\delta([L_0, L_m]) &= [L_0 \otimes 1 + 1 \otimes L_0, \delta(L_m)] - [L_m \otimes 1 + 1 \otimes L_m, \delta(L_0)] \nonumber \\
&=  (-m) \delta(L_m) - [L_m \otimes 1 + 1 \otimes L_m, \delta(L_0)]  \\
\Leftrightarrow 0 &=  [L_m \otimes 1 + 1 \otimes L_m, \delta(L_0)] 
\end{align}
implie that all degree $0$ cocycles vanish on $L_0$ because it has to hold for all $m$ and there is no ad-invariant element in $\bigwedge^2 \left( \mathfrak{W} \oplus \mathfrak{W}\right)$.

As a next step we show that all cocycles are cohomolog to $0$ on $L_{\pm 1}$. Let us assume without loss of generality that the indices of 
\begin{align}
\delta (L_1) = \sum_{i_1 \in I_1} \gamma^1_{i_1} L_{1-i_1} \wedge L_{i_1}
\end{align}
are given by $i_1 \in I_1 = \{-p_1, -p_2, ..., -p_n\vert p_1 > p_2 > ... > p_n > 1, n \in \mathbb{N}\}$. From the cocycle condition we get
\begin{align}
\delta([L_1, L_{-1}]) &= [L_1 \otimes 1 + 1 \otimes L_1, \delta (L_{-1})] - [L_{-1} \otimes 1 + 1 \otimes L_{-1}, \delta (L_{1})]  \\
\Leftrightarrow 0 &= \sum_{i_{-1} \in I_{-1}} (\gamma^{-1}_{i_{-1}}(2 + i_{-1}) L_{-i_{-1}} \wedge L_{i_{-1}} + \gamma^{-1}_{i_{-1}}(i_{-1}-1) L_{1 + i_{-1}}\wedge L_{-1-i_{-1}}) \nonumber \\
& \phantom{=}  + \sum_{j=1}^n (\gamma^1_{-p_j} (2+ p_j) L_{p_j} \wedge L_{-p_j} + \gamma^1_{-p_j}(p_j-1) L_{-1- p_j} \wedge L_{1+ p_j}). \label{p}
\end{align}
Lets focus on the first term in the second line of \eqref{p} with $p_1$; it can only be cancelled by any of the other $p_j$ terms if $p_2= p_1 -1$ which we discuss below. In the case $p_2 \neq p_1-1$ there are two terms that can contribute, one from the first and the second summand in the first line in \eqref{p} which we will call type $I$ and type $II$ terms respectively \footnote{Here and in the following we use the index restriction. Otherwise also e.g. a type $I$ term with $i_{-1} = p_j$ could be used.}. The type $II$ term would correspond to  $i_{-1} = -1-p$. If it existed with non-zero $\gamma^{-1}_{-1-p}$ it would imply the existence of a type $I$ term of the form $\gamma^{-1}_{-1-p} (1-p_1) L_{1+p_1} \wedge L_{-1-p_1}$ which in turn can only be cancelled by a type $II$ term with $i_{-1} = -2-p_1$. Since also none of the prefactors $(2+i_{-1})$ and $(i_{-1} -1)$ vanishes if $p_1 \neq 1$ this would go on forever so that we need infinitely many terms in  $\delta(L_{-1})$ which is not possible. Thus $\gamma^{-1}_{-1-p}=0$ and we need a type $I$ term with $i_{-1} = -p_1$ 
\begin{align}
\gamma^{-1}_{1-p_1} (2 + p_1) L_{p_1 } \wedge L_{-p_1}
\end{align}
which implies a type $II$ term with the same $i_{-1}$
\begin{align}
\gamma^{-1}_{-p_1} (-1-p_1) L_{1-p_1} \wedge L_{p_1 -1}.
\end{align}
This term can be cancelled only by a type $I$ term with $i_{-1} = 1-p_1$ 
\begin{align}
\gamma^{-1}_{1-p_1} (3 - p_1) L_{1-p_1} \wedge L_{p_1-1}
\end{align}
and the corresponding type $II$ term
\begin{align}\label{II1p}
\gamma^{-1}_{1-p_1} p_1 L_{2-p_1} \wedge L_{p_1}
\end{align}
 requires again a type $I$ term with $i_{-1} = -p_1$
\begin{align}\label{Ip}
\gamma^{-1}_{-p_1} (2-p_1) L_{2-p_1} \wedge L_{p_1}
\end{align}
 ending the sequence. The cancellation of \eqref{II1p} with \eqref{Ip} implies the following ratio of the coefficients
\begin{align}
\frac{\gamma^{-1}_{-p_1}}{\gamma^{-1}_{1-p_1} } = \frac{2-p_1}{p_1}
\end{align}
and when considering the 0-cochain
\begin{align}\label{cochain}
r = \gamma_{s} L_{-s} \wedge L_s
\end{align}
with $s = 1-p_1$, implying
\begin{align}
(\partial_0 r)(L_{-1}) \equiv \delta_r(L_{-1}) = \gamma_{1-p_1}( (-p_1) L_{-2 + p_1} \wedge L_{1-p_1} + (-2 + p_1) L_{p_1-1} \wedge L_{-p_1},
\end{align}
we find the same ratio between the two summands. Thus setting $\gamma_{1-p_1} (p_1)= \gamma^{-1}_{1-p_1}$ in the cocycle
\begin{align}
\delta' = \delta + \delta_r
\end{align}
both coefficients ${\gamma'}^{-1}_{-p_1}, {\gamma'}^{-1}_{1-p_1}$ vanish and therefore also ${\gamma'}^1_{p_1}$ has to be zero. 

Next, we have to consider the case $p_2 = p_1 -1$. In \eqref{p} the term 
\begin{align}\label{p1p2}
\gamma^1_{-p_1}(p_1 -1) L_{1+p_1} \wedge L_{-1-p_1}
\end{align}
can be cancelled by a type $I$ term with $i_{-1} = 1 - p_1$ or a type $II$ term with $i_{-1} = -2-p_1$. If the second term does not vanish it implies the existence of a type $I$ term with the same $i_{-1}$ which can only be eliminated by a type $II$ term with $i_{-1} = -3-p_1$ and so on, so that infinitely many terms are necessary, ruling out this option. Using the same cochain as above in \eqref{cochain} with the same choice for $s$ and $\gamma_s$ we can eliminate the coefficient ${\gamma'}^{-1}_{1-p_1}$ and thus the possibility to cancel \eqref{p1p2} with a type $I$ term is not possible which means that ${\gamma'}^{1}_{-p_1}$ has to vanish. 

For the rest of the $p_j, j > 1$ we can iteratively use the same argumentation. In particular the arguments with the infinite number of terms in $\delta(L_{-1})$ can be extended to the higher $j$ as the sequence would stop at the $i_{-1} = -p_{j-1}$ terms which already has to vanish. Furthermore, one has to add coboundaries from the cochains 
\begin{align}
r_j = \gamma_{j} L_{-(1-p_j)} \wedge L_{1-p_j}, \quad j>1
\end{align}
with suitable coefficients $\gamma_{j} (p_j) = {{{\gamma'}^{...}}'}^{-1}_{1-p_j}$ where we define
\begin{align}
{{\delta'}'} = \delta' + \delta_{r_2}, ...
\end{align}
iteratively so that the required terms in ${{\delta'}^{...}}' (L_{-1})$ are eliminated.

Finally, let us explicitely consider the case $p_1 = 1$ that was excluded in the argumentation above. In that case 
\begin{align}
\delta (L_1) = \gamma^1_{-1} L_0 \wedge L_1
\end{align}
and from the cocycle condition we infer that 
\begin{align}
\delta(L_{-1}) = \gamma^1_{-1} L_{-2} \wedge L_{1}.
\end{align}
On $L_{\pm 1}$ $\delta$ then coincides with $\delta_r$, where $r = \gamma^1_{-1}/2 L_1 \wedge L_{-1}$ and thus $\delta' = \delta - \delta_r$  is zero on these elements.
This concludes the proof that $\delta$ is cohomolog to $0$ on $L_{1}$.

%%%
In the next step it will be shown that $\delta(L_{1}) = 0$ implies that $\delta(L_m)=0$ for $m >1$. . Starting from
\begin{align}
\delta (L_2) = \sum_{i_2 \in I_2} \gamma^2_{i_2} L_{2-i_2} \wedge L_{i_2}
\end{align}
one explicitely obtains by using \eqref{coco} with $m = 1, n=2$, $m = 1, n=3$ and $m=1, n=4$
\begin{align}
\delta (L_3) =& - \sum_{i_2 \in I_2} \gamma^2_{i_2}((i_2 -1)L_{3-i_2} \wedge L_{i_2} - (1-i_2) L_{2-i_2}\wedge L_{i_2 +1}) \\
\delta (L_4) =& \sum_{i_2 \in I_2} \frac{\gamma^2_{i_2}}{2} (i_2-1) ((i_2 -2)L_{4-i_2} \wedge L_{i_2} + 2 (1-i_2) L_{3 -i_2} \wedge L_{i_2 + 1} - i_2 L_{2-i_2} \wedge L_{i_2+2}) \\
\delta(L_5) =& - \sum_{i_2 \in I_2} \frac{\gamma^2_{i_2}}{6} (i_2-1)\bigg((i_2 -2)(i_2 -3)L_{5-i_2} \wedge L_{i_2} + 3 (1-i_2) (i_2 -2) L_{4-i_2} \wedge L_{i_2 +1} \nonumber \\
&  + 3(1-i_2) i_2 L_{3 -i_2} \wedge L_{i_2 +2} + i_2 (1+i_2) L_{2-i_2} \wedge L_{i_2 +3} \bigg).
\end{align}
Using the same argumentation as above we can restrict $i_2$ to be bigger than $1$ and we consider the largest index $i'_2$. Then, \eqref{coco} with $m=2, n=3$ yields
\begin{align}
0 = &\sum_{i_2 \in I_2} \bigg( L_{5-i_2} \wedge L_{i_2} \gamma^2_{i_2}\left(\frac{1}{6}(i_2 -1)(i_2-2)(i_2-3) - (i_2 -1)^2 - (i'_2 +1) \right) \nonumber \\
&+ L_{4-i_2} \wedge L_{i_2 +1}\gamma^2_{i_2} \left(\frac{1}{2}(i_2-1)^2(2-i_2 )+(i_2-1) i_2 \right) \nonumber  \\
&+  L_{3-i_2} \wedge L_{i_2 +2}\gamma^2_{i_2} \left( \frac{1}{2}(i_2-1)^2 i_2 + (i_2-1)(i_2-2)  \right) \nonumber \\
&+ L_{2-i_2} \wedge L_{i_2 +3} \left((i_2-1)^2 - (3-i_2) - \frac{(i_2-1)(i_2+1) i_2}{6} \right) \bigg) \label{l2} \\
\Rightarrow \quad 0 &=L_{2-i'_2} \wedge L_{i'_2 +3} \left( (i'_2-1)^2 - (3-i'_2) - \frac{(i'_2-1)(i'_2+1) i'_2}{6} \right)  \label{l3}
\end{align}
and \eqref{l3} implies for $i'_2 > 1, \gamma^2_{i'_2} \neq 0$ the solutions $i'_2 = 3, 4$. $i_2$ can therefore only take the values $i_2 = 2, 3,4$ and one can calculate explicitely that e.g. the term proportional to $L_{1} \wedge L_5$ in \eqref{l2} does not vanish so $\gamma^1_{i_2'} = 0$. Thus $\delta (L_2) = 0$ and iteratively one shows that \eqref{coco} with $m =1$ implies $\delta (L_n) =0$ for $n > 2$. For arbitrary positive $m$ one finds
\begin{align}
\delta([L_{-1}, L_m]) &= - [L_m \otimes 1 + 1 \otimes L_m , \delta(L_{-1})] \\
\Rightarrow 0 &= -\sum_{i_{-1} \in I_{-1}} \gamma^{-1}_{i_{-1}} ((m+1 + i_{-1}) L_{m-1-i_{-1}} \wedge L_{i_{-1}} + (m-i_{-1})L_{-1 - i_{-1}} \wedge L_{i_{-1} +m})
\end{align}
which yields $\gamma^{-1}_{i'_{-1}} =0$ for the largest index $i'_{-1}$ and thus $\delta (L_{-1}) = 0$.

Finally, one shows explicitely that \eqref{coco} with $m=1, n=-2$ results in $\delta (L_{-2}) = 0$ and, similarly to the case of positive $m$ that can be used to show that $\delta (L_m) = 0$ for all $m < -2$, completing the proof of the first theorem.

%%%

%In the next step it will be shown that $\delta(L_{\pm 1}) = 0$ implies that $\delta(L_m) = 0$ for all $m \in \mathbb{Z}$. First, consider $m > 1$. It suffices to show that $\delta(L_2) = 0$ because one can then iteratively use \eqref{coco} with $n = 1$ to show that $\delta(L_m) =0$ for all $m > 2$. Starting from
%\begin{align}
%\delta (L_2) = \sum_{i_2 \in I_2} \gamma^2_{i_2} L_{2-i_2} \wedge L_{i_2}
%\end{align}
%one explicitely obtains by using \eqref{coco} with $m= -1, n = 2$
%\begin{align}\label{l2}
%\sum_{i_2 \in I_2} \gamma^2_{i_2} (i_2 -3) L_{1-i_2} \wedge L_{i_2} + (-1-i_2) L_{2-i_2} \wedge L_{i_2 -1} = 0.
%\end{align}
%Using the same argumentation as above we can restrict $i_2$ to be bigger than $1$ and we consider the largest index $i'_2$. Then the first term in \eqref{l2} can not be cancelled by any other index and if $i'_2 \neq 3$ $\gamma^2_{i'_2}$ has to vanish. If $i'_2 = 3$, because of the restriction on the indices, there could only exist one other index $i_2 = 2$. In that case \eqref{l2} becomes 
%\begin{align}
%\gamma^2_3 (-4) L_{-1} \wedge L_2 + \gamma^2_{2} (- L_{-1} \wedge L_2  - 3 L_0 \wedge L_1) = 0
%\end{align}
%and thus $\gamma^2_{2/3} = 0$. This implies that $\delta(L_2) =0$ and thus also $\delta(L_m) = 0$ for $m>1$. For $m<-1$ the same calculation starting from \eqref{coco} with $m= 1, n = -2$ shows that $\delta(L_m) = 0$, completing the proof of the first theorem.

\subsection{Proof of Theorem 2}

Note that a 1-cocycle $\delta$ applied to an element of $\mathfrak{W}$ can be split into three parts $\delta^I, \delta^{II}, \delta^{III}$, mapping to $\mathfrak{W} \wedge \mathfrak{W}$, $\overline{ \mathfrak{W}} \wedge \overline{\mathfrak{W}}$ or $\mathfrak{W} \wedge \overline{\mathfrak{W}}$ respectively, which have to satisfy the cocycle condition separately. From the previous theorem it follows that $\delta^I$ is cohomolog to zero and from \eqref{coco} one can easily see that $\delta^{II}$ has to vanish. Thus we only need to consider the part $\delta^{III}$ which again can be separated by the degree $d$, which we define such that 
\begin{align}
\delta(L_m) = L_i \wedge \overline{L}_j
\end{align}
has $d = i-m$. 
A general cocycle of homogenous degree $d \neq 0$ is given by
\begin{align}
\delta(L_0) = \sum_{j_0 \in I_0} \alpha^0_{j_0} L_{d} \wedge \overline{L}_{j_0}
\end{align}
on $L_0$. Setting 
\begin{align}
r = \sum_{j_0 \in I_0} \frac{\alpha^0_{j_0}}{d} L_{d} \wedge \overline{L}_{j_0}
\end{align}
we then have 
\begin{align}
\delta'(L_0) = \delta(L_0) - \delta_r(L_0) = 0.
\end{align}
Using this in 
\begin{align}
\delta'([L_0, L_m]) = [L_0 \otimes 1 + 1 \otimes L_0, \delta'(L_m)] -  [L_m \otimes 1 + 1 \otimes L_m, \delta'(L_0)]
\end{align}
it follows that 
\begin{align}
-m \delta'(L_m) = -(d+m) \delta'(L_m) \Rightarrow \delta'(L_m) = 0
\end{align}
concluding the proof for $d\neq 0$.

A general degree $0$ cocycle has the form
\begin{align}
\delta (L_m) = \sum_{i_m \in I_m} \gamma^m_{i_m} L_m \wedge \overline{L}_{i_m}
\end{align}
and by choosing
\begin{align}
r = \sum_{i_1 \in I_1} \gamma_{i_1} L_0 \wedge \overline{L}_{i_1}
\end{align}
it follows that
\begin{align}
\delta'(L_1) = \delta(L_1) - \delta_r(L_1) = 0.
\end{align}
Then for $m\neq 1$
\begin{align}\label{lml1}
\delta'([L_m, L_1]) &= -  [L_1 \otimes 1 + 1 \otimes L_1, \delta'(L_m)] \\
\Rightarrow  \sum_{i_{m+1} \in I_{m+1}} (m-1) \gamma^{m+1}_{i_{m+1}}  L_{m+1} \wedge \overline{L}_{i_{m+1}} &= \sum_{i_m \in I_m} (m-1) \gamma^m_{i_m} L_{m+1} \wedge \overline{L}_{i_m}
\end{align}
and it follows that 
\begin{align}\label{m1}
\gamma^{m+1}_{i_{m+1}} = \gamma^m_{i_m}.
\end{align}
 If $m = 0$ in \eqref{lml1} we conclude 
\begin{align}
0 = - \sum_{i_0 \in I_0} \gamma^0_{i_0} L_1 \wedge \overline{L}_{i_0}
\end{align}
and thus $\gamma^0_{i_0}=0$. Because of \eqref{m1} $\gamma^m_{i_m} = \gamma^0_{i_0}$ for $m<0$ and for $m >0$ all coefficients are given by $\gamma^m_{i_m} = \gamma^2_{i_2}$. However, from \eqref{coco} with $m= 2, n=3$ we find
\begin{align}
- \sum_{i_2 \in I_2} \gamma^2_{i_2} L_5 \wedge \overline{L}_{i_2}= \sum_{i_2 \in I_2} \gamma^2_{i_2} (-2) L_5 \wedge \overline{L}_{i_2} 
\end{align}
and thus $\gamma^2_{i_2} = 0$, concluding the proof.

\section{Classification of Triangular r-matrices} \label{app-b}

First, note that since $\mathfrak{o}(4, \mathbb{C}) = \mathfrak{sl}(2) \oplus  \bar {\mathfrak{sl}}(2) $ and $\bigwedge^2 \mathfrak{o}(4, \mathbb{C}) = \mathfrak{sl}(2) \wedge  {\mathfrak{sl}}(2)  \oplus \mathfrak{sl}(2) \wedge  \bar {\mathfrak{sl}}(2) \oplus \bar {\mathfrak{sl}}(2) \wedge  \bar {\mathfrak{sl}}(2)$  each r-matrix can be split according to 
\begin{align}
r = a + \bar a + b, \quad a \in \mathfrak{sl}(2) \wedge {\mathfrak{sl}}(2),\, \bar a \in \bar{ \mathfrak{sl}}(2) \wedge  \bar {\mathfrak{sl}}(2),\, b \in \mathfrak{sl}(2) \wedge  \bar {\mathfrak{sl}}(2).
\end{align}
Starting with a generic 
\begin{align}\label{ans-a}
a = \alpha_+ L_1 \wedge L_0 + \alpha_0 L_1 \wedge L_{-1} + \alpha_- L_{-1} \wedge L_0
\end{align}
triangularity $[[a, a]] = 0$ enforces 
\begin{align}\label{alin}
\alpha_{0}^2 = - \alpha_+ \alpha_-. 
\end{align}
Using the automorphism \eqref{au1} with $\gamma = \sqrt{- \frac{\alpha_-}{\alpha_+}}, \epsilon = 1$ in the case $\alpha_0 \neq 0$ and with $\epsilon = -1(+1)$ if $\alpha_-=0 (\alpha_+ =0)$ we find that there are two one-parameter r-matrices in $\mathfrak{sl}(2) \wedge  {\mathfrak{sl}}(2)$
\begin{align}
a_1 = \alpha ( L_1 \wedge L_0 +  L_1 \wedge L_{-1} -L_{-1} \wedge L_0), \\
a_2 = \alpha L_1 \wedge L_0
\end{align}
and similar for  $\bar{\mathfrak{sl}}(2) \wedge \bar{\mathfrak{sl}}(2)$
\begin{align}
\bar a_1 = \bar \alpha ( \bar L_1 \wedge \bar L_0 + \bar L_1 \wedge \bar L_{-1} - \bar L_{-1} \wedge \bar L_0), \\
\bar a_2 =   \bar \alpha \bar L_1 \wedge \bar L_0.
\end{align}
For r-matrices that only contain terms of type $b$ one has to demand $[[b, b]] =0$ and the general result (before applying any automorphisms) as obtained in \cite{Borowiec:2015nlw,2015nlw2} reads
\begin{align}
(\beta_+ L_1 + \beta_0 L_0 + \beta_- L_{-1} ) \wedge (\bar \beta_+ \bar L_1 + \bar \beta_0 \bar L_0 + \bar \beta_- \bar L_{-1}).
\end{align}
Taking into account the automorphisms \eqref{au1}, \eqref{au2} one can represent this as eleven r-matrices with up to four parameters 
\begin{align}\label{bb1}
b_1 = & (\beta L_1 + \beta_0 L_0 + \beta L_{-1}) \wedge (\bar \beta \bar L_1 + \bar \beta_0 \bar L_0 + \bar \beta  \bar L_{-1}) ,\\
b_2 = & ( L_1 +L_0 ) \wedge (\bar \beta \bar L_1 + \bar \beta_0 \bar L_0 + \bar \beta  \bar L_{-1}), \\
b_3 = & (L_1 + L_{-1}) \wedge (\bar \beta \bar L_1 + \bar \beta_0 \bar L_0 + \bar \beta  \bar L_{-1}), \\
b_4 = & \beta( L_1 +  L_0 ) \wedge ( \bar L_1 +  \bar L_0 ), \quad b_5 =  \beta( L_1 + L_{-1}) \wedge  (\bar L_1 +  \bar L_{-1} ),  \\
b_6 = & \beta( L_1 + L_{-1}) \wedge  (\bar L_1 +  \bar L_0 ), \quad b_7 = L_1 \wedge  (\bar \beta \bar L_1 + \bar \beta_0 \bar L_0 + \bar \beta  \bar L_{-1}), \\
b_8 = & L_1 \wedge \bar \beta(\bar L_1 + \bar L_0), \quad b_9 = L_1 \wedge \bar \beta(\bar L_1 + \bar L_{-1}), \\
b_{10} = & L_1 \wedge \bar L_1, \quad b_{11} = L_1 \wedge \bar L_0. \label{bb2}
\end{align}
When combining $a$, $\bar a$ and $b$ terms there are two different cases, $[[b, b]] =0$ and $[[b, b]] = -2[[b, a]] - 2[[b, \bar a]] \neq 0$, that will be analysed separately. 
In the first case (for the moment considering only $a$ terms) one infers $[[b, a]] = 0$ and $[[a, a ]] =0$. With the general ansatz \eqref{ans-a} for $a$ and 
\begin{align}
b = & \beta_1 L_{1} \wedge \bar L_{1} +  \beta_2 L_{1} \wedge \bar L_{0} +  \beta_3 L_{1} \wedge \bar L_{-1} \nonumber \\
& +  \beta_4 L_{0} \wedge \bar L_{1} +  \beta_5 L_{0} \wedge \bar L_{0} +  \beta_6 L_{0} \wedge \bar L_{-1} \nonumber \\
& +  \beta_7 L_{-1} \wedge \bar L_{1} +  \beta_8 L_{-1} \wedge \bar L_{0}   \beta_9 L_{-1} \wedge \bar L_{-1} \label{ans-b}
\end{align}
we extract the equations
\begin{align} \label{claseq1}
-2 \beta_1 \alpha_0 + \beta_4 \alpha_1 & = 0, \quad - \beta_4 \alpha_{-1} -2 \beta_7 \alpha_0 =0, \\
\beta_1 \alpha_{-1} + \beta_7 \alpha_1 & = 0, \quad -2 \beta_2 \alpha_0 + \beta_5 \alpha_1 =0 , \\
- \beta_5 \alpha_{-1} -2 \beta_8 \alpha_0 & =0, \quad \beta_2 \alpha_{-1} + \beta_8 \alpha_1 = 0, \\
-2 \beta_3 \alpha_0 + \beta_6  \alpha_1  & =0, \quad -\beta_6 \alpha_{-1} -2 \beta_9 \alpha_0 =0, \\
\beta_3 \alpha_{-1} + \beta_9 \alpha_1 & =0 \label{claseq2}
\end{align}
from $[[b, a]]=0$. For the coefficients of $a$ triangularity entails \eqref{alin} and for $b$ we additionaly use the automorphisms to bring them in the form \eqref{bb1}-\eqref{bb2}. For $b = b_1$, implying 
$$ \beta_1 = \beta_3 = \beta_7 = \beta_9, \quad \beta_2 = \beta_8, \quad \beta_4 = \beta_6, $$
the equations \eqref{claseq1}-\eqref{claseq2} yield
$$ \alpha_{-1} = - \alpha_1, \quad \beta_4 = 2 \beta_1, \quad \beta_5 = 2 \beta_2 $$
resulting in 
\begin{align}
r = (L_1 + L_{-1} + 2 L_0) \wedge ( \beta_1 (\bar L_1 + \bar L_{-1}) + \beta_2 \bar L_0) + a_1.
\end{align}
Similarly for the other r-matrix components of type $b$ one has
\begin{align}\label{a+b1}
r \equiv b_2 + a =& L_1 \wedge (\beta_1 (\bar L_1 + \bar L_{-1}) + \beta_2 \bar L_0) + a_2 , \\
r \equiv b_3+a = & b_3 + \alpha (L_1 - L_{-1}) \wedge L_0, \label{disc1} \\
r \equiv b_4 + a =& \beta L_1 \wedge (\bar L_1 + \bar L_0) + a_2, \\
r \equiv b_4 + a = & \beta (L_1 + L_0) \wedge \bar L_1  + a_2, \\
r \equiv b_5 + a = & b_5  + \alpha (L_1  - L_{-1}) \wedge L_0, \\
r \equiv b_6 + a = & b_6 + a_1,  \label{a+b2} \\
r \equiv b_7 + a = & L_1 \wedge \bar L_0 + a_2, \\
r \equiv b_8 + a = & L_1 \wedge (\bar L_1 + \bar L_0) + a_2, \\
r \equiv b_9 + a = & L_1 \wedge (\bar L_1 + \bar L_{-1}) + a_2, \\
r \equiv b_{10} + a = & L_1 \wedge \bar L_1 + a_2.
\end{align}
To classify r-matrices of the form $b + \bar a$ one can use \eqref{au2} and that the coefficients of $\bar b$ with \eqref{ans-b} are just the transposed coefficients (if they are represented by a $3 \times 3$ matrix) of $b$ and a global minus sign. In the symmetric cases $b_1, b_4, b_5, b_{10}$ the results are automorphic to \eqref{a+b1}-\eqref{a+b2} with $\varphi'$ and for the rest one has
\begin{align}
r =- \bar b_2 + a = & \beta (L_1 + L_{-1} + 2 L_0) \wedge (\bar L_1 + \bar L_0) + a_1, \\
r = - \bar b_4 + a = & \beta (L_1 + L_{-1} + 2 L_0) \wedge (\bar L_1 + \bar L_{-1}) +a_1, \\
r = - \bar b_6 + a  = & b_6 + \alpha L_1 \wedge( L_0 +2 L_{-1}), \label{disc2} \\
r = - \bar b_7 + a = & L_1 \wedge (\bar L_1 + \bar L_{-1} +2 \bar L_0)+ \bar a_1.
\end{align}
While the $r$ in \eqref{disc1} and \eqref{disc2} are solutions of $[[b, a]] = 0$ the $a$ part is not triangular so they have to be discarded.
Combining the previous results (and explicitely calculating some ''overlaps'' of the form $[[b, \bar a]]$) we find for $b + a + \bar a $ the following possibilities
\begin{align}
r \equiv & b_1 + a + \bar a = (L_1 + L_{-1} + 2 L_0) \wedge ( \bar L_1 + \bar L_{1} + 2 \bar L_0) + a_1 + \bar a_1, \\
%r \equiv & b_2 + a + \bar a = L_{1} \wedge \bar L_0  + a_2 + \bar a_2, \\
r \equiv & b_2 + a + \bar a = \beta L_1 \wedge (\bar L_1 + \bar L_{-1} + 2 \bar L_0)  + a_2 + \bar a_1, \\
r \equiv & b_4 + a + \bar a = L_1 \wedge (\bar L_1 + \bar L_0) + a_2 + \bar a_2, \\
r \equiv & b_{10} + a + \bar a = L_1 \wedge \bar L_{1} + a_2 + \bar a_2.
\end{align}

In the case $[[b, b]] \neq 0$ we again make use of the results found in \cite{Borowiec:2015nlw,2015nlw2}. In particular the general solution for the equation 
\begin{align}\label{abgl}
0 \neq [[b, b]] = - 2[[b, a]] - 2[[b, \bar a]]
\end{align}
 up to $\text{Aut}(\mathfrak{o}(4, \mathbb{C}))$ has the form
\begin{align}\label{absol}
\alpha L_1 \wedge L_{-1} - \alpha \bar L_1 \wedge \bar L_{-1} + b, \quad \alpha  L_1 \wedge L_0 + \alpha \bar L_1 \wedge \bar L_0 + b', 
\end{align}
with specific $b, b'$ that are not of interest for now. The first r-matrix in \eqref{absol} is quasitriangular with ad-invariant (in $\mathfrak{o}(4, \mathbb{C})$) element containing $\Omega = 4 \alpha^2 L_1 \wedge L_0 \wedge L_{-1} + ...$ and since the solutions of \eqref{abgl} up to $\text{Aut}(\mathfrak{W} \oplus \mathfrak{W})$ are in the orbits of $\mathfrak{o}(4, \mathbb{C})$ automorphisms $\varphi$ containing \eqref{absol} we would need 
\begin{align}
\varphi (\Omega) = 4 \alpha^2 \varphi (  L_1 \wedge L_0 \wedge L_{-1} ) + ... = 0
\end{align}
to obtain a triangular solution. This, however, would entail that the matrix of the coefficients of $\varphi$ has determinant zero but then it would not be invertible and thus $\varphi$ no automorphism. Furthermore there can be no $\mathfrak{o}(4, \mathbb{C})$ automorphism that maps the $a$ terms of the second solution of \eqref{absol} to $a_1$ because $a_1$ can not be written in the form $ (\alpha_1 L_1 + \alpha_2 L_0 + \alpha_3 L_{-1}) \wedge (\alpha'_1 L_1 + \alpha'_2 L_0 + \alpha'_3 L_{-1}) $. We conclude that only r-matrices of the form $b + a_2 + \bar a_2$ have to be considered. To this end we extract the equations 
\begin{align}
\beta_1 \beta_5 - \beta_2 \beta_4 + \beta_4 2 \alpha & = 0, \quad - \beta_4 \beta_8  + \beta_7 \beta_5 =0, \\
\beta_1 \beta_8 - \beta_2 \beta_7 + \beta_7 2 \alpha & =0, \quad 2 \beta_1 \beta_6 -2 \beta_3 \beta_4 + \beta_5 2 \alpha = 0, \\
-2 \beta_4 \beta_9 + 2 \beta_6 \beta_7 & =0, \quad 2 \beta_1 \beta_9 -2 \beta_3 \beta_7 + \beta_8 2 \alpha =0, \\
- \beta_3 \beta_5 + \beta_2 \beta_6 + \beta_6 2 \alpha & =0, \quad - \beta_5 \beta_9 + \beta_6 \beta_8 =0, \\
- \beta_3 \beta_8 + \beta_2 \beta_9 + \beta_9 2 \alpha & = 0
\end{align}
from \eqref{abgl}. Additionaly we also get the same eqations with $\alpha \rightarrow - \bar \alpha$ and in the terms proportional to $\bar \alpha$ the coefficients of $b$ are transposed. Solving these equations yields only the solution
\begin{align}
\beta_1 L_1 \wedge \bar L_1 + \beta_2( L_1 \wedge \bar L_0 + \bar L_1 \wedge L_0) + \beta_2 L_1 \wedge L_0 + \beta_2 \bar L_1 \wedge \bar L_0, 
%\beta_1 L_1 \wedge \bar L_0 + \beta_2 L_0 \wedge \bar L_1 - \beta_1 L_1 \wedge L_0 - \beta_2 \bar L_1 \wedge \bar L_0.
\end{align}
 i.e. the same as in \cite{Borowiec:2015nlw,2015nlw2}.

After removing duplicacies all the r-matrices we found can be casted into the classes \eqref{classa}-\eqref{classb}.


\begin{thebibliography}{99}

  %\cite{ch:I-1Staruszkiewicz:1963zza}
\bibitem{ch:I-1Staruszkiewicz:1963zza}
  A.~Staruszkiewicz,
  ``Gravitation Theory in Three-Dimensional Space,''
  Acta Phys.\ Polon.\  {\bf 24} (1963) 735.
  %%CITATION = APPOA,24,735;%%
  %163 citations counted in INSPIRE as of 31 Mar 2015
  
  %\cite{Deser:1983tn}
\bibitem{Deser:1983tn}
S.~Deser, R.~Jackiw and G.~'t Hooft,
``Three-Dimensional Einstein Gravity: Dynamics of Flat Space,''
Annals Phys. \textbf{152} (1984), 220
%doi:10.1016/0003-4916(84)90085-X
%1035 citations counted in INSPIRE as of 25 Apr 2021

%\cite{Deser:1983nh}
\bibitem{Deser:1983nh}
S.~Deser and R.~Jackiw,
``Three-Dimensional Cosmological Gravity: Dynamics of Constant Curvature,''
Annals Phys. \textbf{153} (1984), 405-416
%doi:10.1016/0003-4916(84)90025-3
%413 citations counted in INSPIRE as of 25 Apr 2021
  
    %\cite{ch:I-1Carlip:1998uc}
\bibitem{ch:I-1Carlip:1998uc}
  S.~Carlip,
  ``Quantum gravity in 2+1 dimensions,''
  Cambridge, UK: Univ. Pr. (1998) 276 p
  %3 citations counted in INSPIRE as of 31 mar 2015
  
  %\cite{ch:I-2Witten:1988hc}
\bibitem{ch:I-2Witten:1988hc}
  E.~Witten,
  ``(2+1)-Dimensional Gravity as an Exactly Soluble System,''
  Nucl.\ Phys.\  B {\bf 311} (1988) 46.
  %%CITATION = NUPHA,B311,46;%%

  %\cite{ch:I-2Achucarro:1987vz}
\bibitem{ch:I-2Achucarro:1987vz}
  A.~Achucarro and P.~K.~Townsend,
  ``A Chern-Simons Action for Three-Dimensional anti-De Sitter Supergravity Theories,''
  Phys.\ Lett.\ B {\bf 180} (1986) 89.
  %%CITATION = PHLTA,B180,89;%%
  
  %\cite{Brown:1986nw}
\bibitem{Brown:1986nw}
J.~D.~Brown and M.~Henneaux,
``Central Charges in the Canonical Realization of Asymptotic Symmetries: An Example from Three-Dimensional Gravity,''
Commun. Math. Phys. \textbf{104} (1986), 207-226
%doi:10.1007/BF01211590
%1959 citations counted in INSPIRE as of 25 Apr 2021

%\cite{Maldacena:1997re}
\bibitem{Maldacena:1997re}
J.~M.~Maldacena,
``The Large N limit of superconformal field theories and supergravity,''
Adv. Theor. Math. Phys. \textbf{2} (1998), 231-252
%doi:10.1023/A:1026654312961
[arXiv:hep-th/9711200 [hep-th]].
%16571 citations counted in INSPIRE as of 25 Apr 2021

%\cite{Kraus:2006wn}
\bibitem{Kraus:2006wn}
P.~Kraus,
``Lectures on black holes and the AdS(3) / CFT(2) correspondence,''
Lect. Notes Phys. \textbf{755} (2008), 193-247
[arXiv:hep-th/0609074 [hep-th]].
%244 citations counted in INSPIRE as of 25 Apr 2021

%\cite{Freidel:2008sh}
\bibitem{Freidel:2008sh}
L.~Freidel,
``Reconstructing AdS/CFT,''
[arXiv:0804.0632 [hep-th]].
%49 citations counted in INSPIRE as of 25 Apr 2021

%\cite{Banados:1992wn}
\bibitem{Banados:1992wn}
M.~Banados, C.~Teitelboim and J.~Zanelli,
``The Black hole in three-dimensional space-time,''
Phys. Rev. Lett. \textbf{69} (1992), 1849-1851
%doi:10.1103/PhysRevLett.69.1849
[arXiv:hep-th/9204099 [hep-th]].
%2825 citations counted in INSPIRE as of 25 Apr 2021

%\cite{Banados:1992gq}
\bibitem{Banados:1992gq}
M.~Banados, M.~Henneaux, C.~Teitelboim and J.~Zanelli,
``Geometry of the (2+1) black hole,''
Phys. Rev. D \textbf{48} (1993), 1506-1525
[erratum: Phys. Rev. D \textbf{88} (2013), 069902]
%doi:10.1103/PhysRevD.48.1506
[arXiv:gr-qc/9302012 [gr-qc]].
%1600 citations counted in INSPIRE as of 25 Apr 2021

%\cite{Carlip:2005zn}
\bibitem{Carlip:2005zn}
S.~Carlip,
``Conformal field theory, (2+1)-dimensional gravity, and the BTZ black hole,''
Class. Quant. Grav. \textbf{22} (2005), R85-R124
%doi:10.1088/0264-9381/22/12/R01
[arXiv:gr-qc/0503022 [gr-qc]].
%246 citations counted in INSPIRE as of 25 Apr 2021

%\cite{Witten:1988hf}
\bibitem{Witten:1988hf}
E.~Witten,
``Quantum Field Theory and the Jones Polynomial,''
Commun. Math. Phys. \textbf{121} (1989), 351-399
%doi:10.1007/BF01217730
%3086 citations counted in INSPIRE as of 26 Apr 2021

%\cite{Fock:1998nu}
\bibitem{Fock:1998nu}
V.~V.~Fock and A.~A.~Rosly,
``Poisson structure on moduli of flat connections on Riemann surfaces and r matrix,''
Am. Math. Soc. Transl. \textbf{191} (1999), 67-86
[arXiv:math/9802054 [math.QA]].
%96 citations counted in INSPIRE as of 26 Apr 2021

%\cite{Alekseev:1993qs}
\bibitem{Alekseev:1993qs}
A.~Y.~Alekseev and A.~Z.~Malkin,
``Symplectic structures associated to Lie-Poisson groups,''
Commun. Math. Phys. \textbf{162} (1994), 147-174
%doi:10.1007/BF02105190
[arXiv:hep-th/9303038 [hep-th]].
%107 citations counted in INSPIRE as of 26 Apr 2021

%\cite{Alekseev:1993rj}
\bibitem{Alekseev:1993rj}
A.~Y.~Alekseev and A.~Z.~Malkin,
``Symplectic structure of the moduli space of flat connection on a Riemann surface,''
Commun. Math. Phys. \textbf{169} (1995), 99-120
%doi:10.1007/BF02101598
[arXiv:hep-th/9312004 [hep-th]].
%87 citations counted in INSPIRE as of 26 Apr 2021

%\cite{Bais:2002ye}
\bibitem{Bais:2002ye}
F.~A.~Bais, N.~M.~Muller and B.~J.~Schroers,
``Quantum group symmetry and particle scattering in (2+1)-dimensional quantum gravity,''
Nucl. Phys. B \textbf{640} (2002), 3-45
%doi:10.1016/S0550-3213(02)00572-2
[arXiv:hep-th/0205021 [hep-th]].
%62 citations counted in INSPIRE as of 26 Apr 2021

%\cite{Meusburger:2003ta}
\bibitem{Meusburger:2003ta}
C.~Meusburger and B.~J.~Schroers,
``Poisson structure and symmetry in the Chern-Simons formulation of (2+1)-dimensional gravity,''
Class. Quant. Grav. \textbf{20} (2003), 2193-2234
%doi:10.1088/0264-9381/20/11/318
[arXiv:gr-qc/0301108 [gr-qc]].
%61 citations counted in INSPIRE as of 26 Apr 2021

%\cite{Meusburger:2003hc}
\bibitem{Meusburger:2003hc}
C.~Meusburger and B.~J.~Schroers,
``The quantisation of Poisson structures arising inChern-Simons theory with gauge group $G \ltimes \mathfrak{g}^*$,''
Adv. Theor. Math. Phys. \textbf{7} (2003) no.6, 1003-1043
%doi:10.4310/ATMP.2003.v7.n6.a3
[arXiv:hep-th/0310218 [hep-th]].
%50 citations counted in INSPIRE as of 26 Apr 2021

%\cite{Meusburger:2007ad}
\bibitem{Meusburger:2007ad}
C.~Meusburger and B.~J.~Schroers,
``Quaternionic and Poisson-Lie structures in 3d gravity: The Cosmological constant as deformation parameter,''
J. Math. Phys. \textbf{49} (2008), 083510
%doi:10.1063/1.2973040
[arXiv:0708.1507 [gr-qc]].
%36 citations counted in INSPIRE as of 26 Apr 2021

%\cite{Freidel:2005bb}
\bibitem{Freidel:2005bb}
L.~Freidel and E.~R.~Livine,
``Ponzano-Regge model revisited III: Feynman diagrams and effective field theory,''
Class. Quant. Grav. \textbf{23} (2006), 2021-2062
%doi:10.1088/0264-9381/23/6/012
[arXiv:hep-th/0502106 [hep-th]].
%208 citations counted in INSPIRE as of 26 Apr 2021

%\cite{Freidel:2005me}
\bibitem{Freidel:2005me}
L.~Freidel and E.~R.~Livine,
``3D Quantum Gravity and Effective Noncommutative Quantum Field Theory,''
Phys. Rev. Lett. \textbf{96} (2006), 221301
%doi:10.1103/PhysRevLett.96.221301
[arXiv:hep-th/0512113 [hep-th]].
%243 citations counted in INSPIRE as of 26 Apr 2021

%\cite{Bondi:1962px}
\bibitem{Bondi:1962px}
  H.~Bondi, M.~G.~J.~van der Burg and A.~W.~K.~Metzner,
  ``Gravitational waves in general relativity. 7. Waves from axisymmetric isolated systems,''
  Proc.\ Roy.\ Soc.\ Lond.\ A {\bf 269} (1962) 21.
  %doi:10.1098/rspa.1962.0161
  %%CITATION = doi:10.1098/rspa.1962.0161;%%
  %933 citations counted in INSPIRE as of 26 Oct 2018
  
  %\cite{Sachs:1962wk}
\bibitem{Sachs:1962wk}
R.~K.~Sachs,
``Gravitational waves in general relativity. 8. Waves in asymptotically flat space-times,''
Proc. Roy. Soc. Lond. A \textbf{270} (1962), 103-126
%doi:10.1098/rspa.1962.0206
%1013 citations counted in INSPIRE as of 26 Apr 2021

%\cite{Sachs:1962zza}
\bibitem{Sachs:1962zza}
R.~Sachs,
``Asymptotic symmetries in gravitational theory,''
Phys. Rev. \textbf{128} (1962), 2851-2864
%doi:10.1103/PhysRev.128.2851
%490 citations counted in INSPIRE as of 26 Apr 2021

%\cite{Borowiec:2018rbr}
\bibitem{Borowiec:2018rbr}
A.~Borowiec, L.~Brocki, J.~Kowalski-Glikman and J.~Unger,
``$\kappa$-deformed BMS symmetry,''
Phys. Lett. B \textbf{790} (2019), 415-420
%doi:10.1016/j.physletb.2019.01.063
[arXiv:1811.05360 [hep-th]].
%2 citations counted in INSPIRE as of 26 Apr 2021

%\cite{Borowiec:2020ddg}
\bibitem{Borowiec:2020ddg}
A.~Borowiec, L.~Brocki, J.~Kowalski-Glikman and J.~Unger,
``BMS algebras in 4 and 3 dimensions, their quantum deformations and duals,''
JHEP \textbf{02} (2021), 084
%doi:10.1007/JHEP02(2021)084
[arXiv:2010.10224 [hep-th]].
%1 citations counted in INSPIRE as of 12 Mar 2021

%\cite{Campiglia:2014yka}
\bibitem{Campiglia:2014yka}
M.~Campiglia and A.~Laddha,
%``Asymptotic symmetries and subleading soft graviton theorem,''
Phys. Rev. D \textbf{90} (2014) no.12, 124028
doi:10.1103/PhysRevD.90.124028
[arXiv:1408.2228 [hep-th]].
%182 citations counted in INSPIRE as of 04 Sep 2021

%\cite{Cianfrani:2016ogm}
\bibitem{Cianfrani:2016ogm}
F.~Cianfrani, J.~Kowalski-Glikman, D.~Pranzetti and G.~Rosati,
``Symmetries of quantum spacetime in three dimensions,''
Phys. Rev. D \textbf{94} (2016) no.8, 084044
%doi:10.1103/PhysRevD.94.084044
[arXiv:1606.03085 [hep-th]].
%32 citations counted in INSPIRE as of 26 Apr 2021

%\cite{ch:II-1Lukierski:1991pn}
\bibitem{ch:II-1Lukierski:1991pn}
  J.~Lukierski, H.~Ruegg, A.~Nowicki and V.~N.~Tolstoi,
  ``Q deformation of Poincare algebra,''
  Phys.\ Lett.\ B {\bf 264}, 331 (1991).

%\cite{ch:II-1Lukierski:1992dt}
\bibitem{ch:II-1Lukierski:1992dt}
  J.~Lukierski, A.~Nowicki and H.~Ruegg,
  ``New quantum Poincare algebra and k deformed field theory,''
  Phys.\ Lett.\ B {\bf 293}, 344 (1992).
  
  %\cite{ch:II-1Lukierski:1993wx}
\bibitem{ch:II-1Lukierski:1993wx}
  J.~Lukierski, H.~Ruegg and W.~J.~Zakrzewski,
  ``Classical quantum mechanics of free kappa relativistic systems,''
{\it  Annals Phys.}\  {\bf 243} 90 (1995)
  [hep-th/9312153].
  %%CITATION = HEP-TH/9312153;%%
  
%\cite{ch:II-1Majid:1994cy}
\bibitem{ch:II-1Majid:1994cy}
  S.~Majid and H.~Ruegg,
  ``Bicrossproduct structure of kappa Poincare group and noncommutative geometry,''
Phys.\ Lett.\ B {\bf 334}, 348 (1994)
%doi:10.1016/0370-2693(94)90699-8
[hep-th/9405107].  


  



%\cite{Borowiec:2017apk}
\bibitem{Borowiec:2017apk}
A.~Borowiec, J.~Lukierski and V.~N.~Tolstoy,
``Basic quantizations of $D=4$ Euclidean, Lorentz, Kleinian and quaternionic $\mathfrak{o}^{\star}(4)$ symmetries,''
JHEP \textbf{11} (2017), 187
%doi:10.1007/JHEP11(2017)187
[arXiv:1708.09848 [hep-th]].
%5 citations counted in INSPIRE as of 25 Jan 2021

%\cite{Ng:2000}
\bibitem{Ng:2000}
Siu-Hung~Ng, Earl~J.~Taft,
''Classification of the Lie bialgebra structures on the Witt and Virasoro algebras'',
Journal of Pure and Applied Algebra,
Volume 151, Issue 1,
2000, Pages 67-88, ISSN 0022-4049.
%https://doi.org/10.1016/S0022-4049(99)00045-6.
%(https://www.sciencedirect.com/science/article/pii/S0022404999000456)

%\cite{Borowiec:2015nlw}
 \bibitem{Borowiec:2015nlw}
A.~Borowiec, J.~Lukierski and V.~N.~Tolstoy,
``Quantum deformations of $D =  4$ Euclidean, Lorentz, Kleinian and quaternionic $\mathfrak{o}^*(4)$ symmetries in unified $\mathfrak{o}(4;\mathbb{C})$ setting,''
Phys. Lett. B \textbf{754} (2016), 176-181
%doi:10.1016/j.physletb.2016.01.016
[arXiv:1511.03653 [hep-th]]. 
\bibitem{2015nlw2}
A.~Borowiec, J.~Lukierski and V.~N.~Tolstoy,
Addendum: to “Quantum deformations of $D$= 4 Euclidean, Lorentz, Kleinian and quaternionic $\mathfrak{o}^*(4)$ symmetries in unified $\mathfrak{o}(4;\mathbb{C})$ setting”
Phys. Lett. B \textbf{770} (2017), 426-430 [arXiv:1704.06852[hep-th]].
%doi:10.1016/j.physletb.2017.04.070
%17 citations counted in INSPIRE as of 05 May 2021

  %\cite{Strominger:2017zoo}
\bibitem{Strominger:2017zoo}
A.~Strominger,
``Lectures on the Infrared Structure of Gravity and Gauge Theory,''
[arXiv:1703.05448 [hep-th]].
%340 citations counted in INSPIRE as of 09 Oct 2020

%\cite{Kowalski-Glikman:2019ttm}
\bibitem{Kowalski-Glikman:2019ttm}
J.~Kowalski-Glikman, J.~Lukierski and T.~Trze\'sniewski,
``Quantum D = 3 Euclidean and Poincar\'e symmetries from contraction limits,''
JHEP \textbf{09} (2020), 096
%doi:10.1007/JHEP09(2020)096
[arXiv:1911.09538 [hep-th]].
%2 citations counted in INSPIRE as of 25 Jan 2021

%\cite{Safari:2019zmc}
\bibitem{Safari:2019zmc}
H.~R.~Safari and M.~M.~Sheikh-Jabbari,
``BMS$_{4}$ algebra, its stability and deformations,''
JHEP \textbf{04} (2019), 068
%doi:10.1007/JHEP04(2019)068
[arXiv:1902.03260 [hep-th]].
%17 citations counted in INSPIRE as of 30 Apr 2021

%\cite{Enriquez-Rojo:2021rtv}
\bibitem{Enriquez-Rojo:2021rtv}
M.~Enriquez-Rojo, T.~Proch\'azka and I.~Sachs,
``On deformations and extensions of $\text{Diff}(S^2)$,''
[arXiv:2105.13375 [hep-th]].
%0 citations counted in INSPIRE as of 27 Sep 2021

%\cite{Compere:2019bua}
\bibitem{Compere:2019bua}
G.~Comp\`ere, A.~Fiorucci and R.~Ruzziconi,
``The $\Lambda$-BMS$_4$ group of dS$_4$ and new boundary conditions for AdS$_4$,''
Class. Quant. Grav. \textbf{36} (2019) no.19, 195017
%doi:10.1088/1361-6382/ab3d4b
[arXiv:1905.00971 [gr-qc]].
%30 citations counted in INSPIRE as of 12 Mar 2021

%\cite{Compere:2020lrt}
\bibitem{Compere:2020lrt}
G.~Comp\`ere, A.~Fiorucci and R.~Ruzziconi,
``The $\Lambda$-BMS$_4$ charge algebra,''
JHEP \textbf{10} (2020), 205
%doi:10.1007/JHEP10(2020)205
[arXiv:2004.10769 [hep-th]].
%27 citations counted in INSPIRE as of 27 Apr 2021

%\cite{Oblak:2016eij}
\bibitem{Oblak:2016eij}
B.~Oblak,
``BMS Particles in Three Dimensions,''
%doi:10.1007/978-3-319-61878-4
[arXiv:1610.08526 [hep-th]].
%80 citations counted in INSPIRE as of 30 Apr 2021

%\cite{Compere:2018aar}
\bibitem{Compere:2018aar}
G.~Comp\`ere and A.~Fiorucci,
``Advanced Lectures on General Relativity,''
[arXiv:1801.07064 [hep-th]].
%73 citations counted in INSPIRE as of 30 Apr 2021



%\cite{chari1995guide}
\bibitem{chari1995guide}
Chari, V. and Pressley, A.,
 A Guide to Quantum Groups,
ISBN 9780521558846,
Cambridge University Press

%\cite{klimyk2011quantum}
\bibitem{klimyk2011quantum}
Klimyk, A. and Schm{\"u}dgen, K.
Quantum Groups and Their Representations,
ISBN 9783642646010,
Springer Berlin Heidelberg

%\cite{etingof2010lectures}
\bibitem{etingof2010lectures}
Etingof, P.I. and Schiffmann, O.
Lectures on Quantum Groups,
ISBN 9781571462077,
International Press


%\cite{Ecker:2019thw}
\bibitem{Ecker:2019thw}
J.~Ecker and M.~Schlichenmaier,
``The low-dimensional algebraic cohomology of the Witt and the Virasoro algebra,''
J. Phys. Conf. Ser. \textbf{1194} (2019) no.1, 012032
%doi:10.1088/1742-6596/1194/1/012032
%0 citations counted in INSPIRE as of 12 Mar 2021

%\cite{Stachura_1998}
\bibitem{Stachura_1998}
Piotr Stachura, 
Poisson-Lie structures on Poincar{\'{e}} and Euclidean groups in three dimensions, 
,1998,
{IOP} Publishing, {31}, {19}, {4555--4564},
{Journal of Physics A: Mathematical and General},

%\cite{Borowiec:2014aqa}
\bibitem{Borowiec:2014aqa}
A.~Borowiec and A.~Pachol,
%``$\kappa$-Deformations and Extended $\kappa$-Minkowski Spacetimes,''
SIGMA \textbf{10} (2014), 107
doi:10.3842/SIGMA.2014.107
[arXiv:1404.2916 [math-ph]].
%22 citations counted in INSPIRE as of 15 Jun 2021

%\cite{Penrose:1976}
\bibitem{Penrose:1976}
R.~Penrose,
``The geometry of impulsive gravitational waves,” General Relativity, Papers in Honour of J. L. Synge (1972): 101-115.

%\cite{Strominger:2016wns}
\bibitem{Strominger:2016wns}
A.~Strominger and A.~Zhiboedov,
``Superrotations and Black Hole Pair Creation,''
Class. Quant. Grav. \textbf{34} (2017) no.6, 064002
%doi:10.1088/1361-6382/aa5b5f
[arXiv:1610.00639 [hep-th]].
%47 citations counted in INSPIRE as of 06 Apr 2021

%\cite{Gleiser:1989vt}
\bibitem{Gleiser:1989vt}
R.~Gleiser and J.~Pullin,
``ARE COSMIC STRINGS GRAVITATIONALLY STABLE TOPOLOGICAL DEFECTS?,''
Class. Quant. Grav. \textbf{6} (1989), L141-L144
%doi:10.1088/0264-9381/6/8/005
%28 citations counted in INSPIRE as of 06 Apr 2021

%\cite{Junbo:2010}
\bibitem{Junbo:2010}
Li.~Junbo , Su.~Yucai and Xin.~Bin. (2010). ``Lie Bialgebra Structures on the Centerless W-Algebra W(2,2)'' Algebra Colloquium. 17. %10.1142/S1005386710000192. 

%\cite{Compere:2013bya}
\bibitem{Compere:2013bya}
G.~Comp\`ere, W.~Song and A.~Strominger,
``New Boundary Conditions for AdS3,''
JHEP \textbf{05} (2013), 152
doi:10.1007/JHEP05(2013)152
[arXiv:1303.2662 [hep-th]].
%117 citations counted in INSPIRE as of 16 Sep 2021

%\cite{Fuentealba:2020zkf}
\bibitem{Fuentealba:2020zkf}
O.~Fuentealba, H.~A.~Gonz\'alez, A.~P\'erez, D.~Tempo and R.~Troncoso,
``Superconformal Bondi-Metzner-Sachs Algebra in Three Dimensions,''
Phys. Rev. Lett. \textbf{126} (2021) no.9, 091602
doi:10.1103/PhysRevLett.126.091602
[arXiv:2011.08197 [hep-th]].
%2 citations counted in INSPIRE as of 16 Sep 2021

%\cite{Afshar:2016wfy}
\bibitem{Afshar:2016wfy}
H.~Afshar, S.~Detournay, D.~Grumiller, W.~Merbis, A.~Perez, D.~Tempo and R.~Troncoso,
``Soft Heisenberg hair on black holes in three dimensions,''
Phys. Rev. D \textbf{93} (2016) no.10, 101503
doi:10.1103/PhysRevD.93.101503
[arXiv:1603.04824 [hep-th]].
%121 citations counted in INSPIRE as of 16 Sep 2021

%\cite{Grumiller:2019fmp}
\bibitem{Grumiller:2019fmp}
D.~Grumiller, A.~P\'erez, M.~M.~Sheikh-Jabbari, R.~Troncoso and C.~Zwikel,
``Spacetime structure near generic horizons and soft hair,''
Phys. Rev. Lett. \textbf{124} (2020) no.4, 041601
doi:10.1103/PhysRevLett.124.041601
[arXiv:1908.09833 [hep-th]].
%48 citations counted in INSPIRE as of 16 Sep 2021

%\cite{Batlle:2020hia}
\bibitem{Batlle:2020hia}
C.~Batlle, V.~Campello and J.~Gomis,
``A canonical realization of the Weyl BMS symmetry,''
Phys. Lett. B \textbf{811} (2020), 135920
%doi:10.1016/j.physletb.2020.135920
[arXiv:2008.10290 [hep-th]].
%2 citations counted in INSPIRE as of 16 Sep 2021

%\cite{Adami:2020ugu}
\bibitem{Adami:2020ugu}
H.~Adami, M.~M.~Sheikh-Jabbari, V.~Taghiloo, H.~Yavartanoo and C.~Zwikel,
``Symmetries at null boundaries: two and three dimensional gravity cases,''
JHEP \textbf{10} (2020), 107
doi:10.1007/JHEP10(2020)107
[arXiv:2007.12759 [hep-th]].
%21 citations counted in INSPIRE as of 16 Sep 2021

\bibitem{Barnich17} 
G. Barnich, "Centrally extended BMS4 Lie algebroid", JHEP 06 (2017) 007
[arXiv:1703.08704].

%\cite{Lu96}
\bibitem{Lu96}
J.H.~Lu,
``Hopf algebroids and quantum groupoids,''
Int. J. Math.7 (1996), 47–70.

%\cite{Brzezinski:2002}
\bibitem{Brzezinski:2002}
T. Brzezinski, G. Militaru,
"Bialgebroids, $\star$ A-Bialgebras and Duality",
Journal of Algebra - J ALGEBRA. 251. 279-294. 

%\cite{Pachol:2017}
\bibitem{Pachol17}
A. Borowiec, A. Pachol,
"Twisted bialgebroids versus bialgebroids from a Drinfeld twist",
 J. Phys. \textbf{A50} (2017) 5, 055205 [1603.09280 [math-ph]].
\end{thebibliography}
\end{document}